\documentclass[a4paper,11pt]{article}
\usepackage{jheppub}

\usepackage{amsfonts}
\usepackage{amssymb}
\usepackage{amsmath}
\usepackage{esint}                   % For closed integrals
\usepackage{epsfig}
\usepackage{latexsym}
\usepackage{color}
\usepackage{graphicx}
\usepackage{hyperref}
\usepackage{nicefrac}
\usepackage[latin1]{inputenc}
\usepackage{pstricks}
\usepackage{slashed}
\usepackage{cancel}
\usepackage{empheq}
\usepackage{multirow}
\def\be{\begin{equation}}
\def\ee{\end{equation}}
\def\ba{\begin{eqnarray}}
\def\ea{\end{eqnarray}}

\usepackage[ddmmyyyy,hhmmss]{datetime}

\def\k{\kappa}

\def\a{\alpha}

\def\b{\beta}

\def\g{\gamma}

\def\D{\Delta}
\def\e{\epsilon}

\def\th{\theta}

\def\m{\mu}

\def\om{\omega}

\def\l{\lambda}

\def\s{\sigma}

\def\cX{{\cal X}}

\def\IR{\relax{\rm I\kern-.18em R}}

\def\inv{^{\raise.0ex\hbox{${\scriptscriptstyle -}$}\kern-.05em 1}}

\def \ov {\over}

\title{Marginally deformed Schr\"{o}dinger/dipole CFT correspondence}

\author{George Georgiou$^{a}$, Georgios Itsios$^{b}$ and Dimitrios Zoakos$^{a,c}$}

\affiliation[a]{Department of Physics, National and Kapodistrian University of Athens, 15784 Athens, Greece}
\affiliation[b]{Institut f\"{u}r Physik, Humboldt-Universit\"{a}t zu Berlin,
IRIS Geb\"{a}ude, Zum Gro\ss en Windkanal 2, 12489 Berlin, Germany}

\affiliation[c]{Department of Engineering and Informatics, 
Hellenic American University, 436 Amherst st, Nashua, NH 03063 USA}

\emailAdd{ggeo@phys.uoa.gr}
\emailAdd{georgios.itsios@physik.hu-berlin.de} \emailAdd{zoakos@gmail.com}

\abstract{We construct and thoroughly study a new integrable example of the AdS/CFT correspondence with Schr\"{o}dinger symmetry.
On the gravity side, the supergravity solution depends on two parameters and is obtained by marginally deforming the internal space of the Schr\"{o}dinger background through a series of TsT transformations. 
On the field theory side, we identify the dual field theory which also depends on two parameters.

We find a point-like string solution and derive its dispersion relation. A non-trivial test of the correspondence is provided by  using the Landau-Lifshitz coherent state approach to reproduce the leading, in  the deformation parameters, terms  of that relation. 
Then, we calculate the Wilson loop, describing the quark/anti-quark  potential at strong coupling. It exhibits confining behaviour when the separation length is much less than the Schr\"{o}dinger parameter.  When the separation length is much greater than the Schr\"{o}dinger parameter the  behaviour is that of a conformal theory.
Subsequently, we take the Penrose limit along a certain null geodesic of the constructed background and calculate the bosonic spectrum. 
Based on that spectrum, we make an educated guess for
the exact, in the 't Hooft coupling,  dispersion relation of the magnon excitations in the original doubly deformed
background. This provides us with an exact prediction for the dimensions of the dual field
theory operators.
}

\begin{document}

\maketitle
\flushbottom

\section{Introduction}

One of the most fascinating discoveries in theoretical physics is the so-called AdS/CFT correspondence \cite{Maldacena:1997re}.
In its original form it asserts the complete equivalence between type-IIB string theory on the $AdS_5 \times S^5$ background and the maximally supersymmetric gauge theory in 4 dimensions, namely $\mathcal {N}=4$ Super Yang-Mills (SYM). The importance of the correspondence relies on its weak/strong coupling nature which provides powerful tools in understanding the strongly coupled dynamics of gauge theories. 
 One of the main properties of the correspondence is its integrability.
 Exploiting the key feature of integrability which both of the aforementioned theories possess, an intense activity took place allowing the determination of its  spectrum, at the planar level, for any value of the 't Hooft coupling $\lambda$. To achieve this goal  a variety of integrability based techniques were employed, such as the asymptotic Bethe ansatz \cite{Staudacher:2004tk}, the thermodynamic Bethe ansatz \cite{Ambjorn:2005wa}, and the Y-system \cite{Gromov:2009tv}.
 
On the contrary, much less is known about the structure constants of the theory. The main obstacle here, is that for the calculation of the three-point correlation functions the exact form of the eigenstates of the dilatation operator is  needed \cite{Georgiou:2008vk,Georgiou:2009tp,Georgiou:2011xj}.
Systematic studies of three-point correlators involving BPS, as well as, non-BPS operators  were performed in \cite{Okuyama:2004bd,Roiban:2004va,Alday:2005nd,Georgiou:2012zj} by computing the contributions arising from both the planar one-loop Feynman diagrams, as well as from the correct form of the one-loop eigenstates of the dilatation operator \cite{Georgiou:2012zj,Georgiou:2009tp}.

More recently, certain non-perturbative in nature methods for bootstrapping three-point correlation functions were developed in \cite{Escobedo:2010xs,Jiang:2014mja,Basso:2015zoa,Kazama:2016cfl}. In addition, by exploiting the AdS/CFT correspondence,  three-point correlation functions involving three {\it heavy} states in the $SU(2)$ or the  $SL(2)$ subsectors were calculated at strong coupling \cite{Kazama:2013qsa,Kazama:2011cp}. This was accomplished by evaluating the area of the corresponding minimal surface through Pohlmeyer reduction. Another front where significant progress has been made is the one where the three-point correlator involves two non-protected operators with large charges, which are dual to classical string solutions, and one light state. The strong coupling result  for this kind of three-point functions can be obtained by integrating the vertex operator that describes the  light state over the classical surface which corresponds to the free propagation of the {\it heavy} state from one point of the boundary of $AdS_5$ to another \cite{Zarembo:2010rr,Costa:2010rz,Roiban:2010fe,Georgiou:2010an,Georgiou:2011qk,Bajnok:2016xxu}.

There is another occasion in which  one can extract useful information for both the spectrum and the structure constants of the theory. This is possible in a particularly interesting limit, called the BMN limit \cite{Berenstein:2002jq} in which one focuses on the sector of operators with large $R$-charge. These operators are dual to string states propagating in the pp-wave limit of the $AdS_5 \times S^5$
geometry.
Different proposals concerning the form of the cubic string Hamiltonian had been put forward  in \cite{Spradlin:2002ar,Pankiewicz:2002tg,DiVecchia:2003yp}.
In \cite{Dobashi:2004nm,Lee:2004cq} the issue of how to correctly relate the string amplitudes obtained from  the pp-wave cubic Hamiltonian to the structure constants of the $\mathcal {N}=4$ SYM was finally settled. This was achieved by combining a number of results available from both  the string and the field theory sides \cite{Georgiou:2004ty,Georgiou:2003kt,Georgiou:2003aa,Chu:2002pd}.\footnote{Furthermore, using the techniques of coordinate and algebraic Bethe ansatz the entanglement entropy of the $\mathcal {N}=4$ SYM spin chain was studied in \cite{Georgiou:2016kge}. An exact expression for the entanglement entropy of a state with  two excitations in the BMN limit was also derived in the same work.} It is precisely this limit that we will employ in section \ref{pp-wave} in order to get information for the spectrum of the marginally deformed  conformal field theory (CFT) with Schrodinger symmetry.

Despite the aforementioned unprecedented insights into the strongly coupled dynamics of gauge theories originating from the original AdS/CFT scenario, $\mathcal {N}=4$ SYM is far from QCD, the theory one would like to understand at low energies where the confinement of the quarks takes place.
In recent years, the identification of integrable deformations of the original AdS/CFT correspondence has attracted significant attention. These come a step closer to more realistic gauge theories. One of the reasons is that due to the deformations supersymmetry may be  completely or partially  broken. A case where the effects of the deformation is more profound is the correspondence between  a certain Schr\"{o}dinger spacetime and its dual null-dipole deformed conformal field theory \cite{Maldacena:2008wh,Herzog:2008wg,Adams:2008wt}. 
On the gravity side \cite{Alishahiha:2003ru} the theory may be easily obtained from the $AdS_5 \times S^5$ background through a solution generating technique known as TsT transformation. One starts by performing an Abelian T-duality along one of the isometries of the five-sphere  $S^5$ followed by a shift along one of the light-like directions of the $AdS_5$ boundary. Finally a second T-duality along the  coordinate of the sphere dualised initially must be performed.
The background resulting from this procedure is the so-called $Sch_5 \times S^5$ and for certain choices of the five sphere isometry is non-supersymmetric.
The holographic dual field theory  is, of course,  also non-supersymmetric and realises the Schr\"{o}dinger symmetry algebra as its symmetry group. 
The field theory dual can be straightforwardly obtained by introducing a certain $\star$-product among the fields of the $\mathcal {N}=4$ SYM Lagrangian. This $\star$-product can be identified with the corresponding Drinfeld--Reshetikhin twist of the underlying integrable structure of the undeformed parent theory, namely $\mathcal {N}=4$ SYM 
\cite{Beisert:2005if,Ahn:2010ws,Matsumoto:2015uja,Kyono:2016jqy,vanTongeren:2015uha}. 
Consequently, the deformed theory is fully integrable at the planar level with its integrability properties being inherited from the parent $\mathcal {N}=4$ SYM.

In contradistinction to the original AdS/CFT scenario very few observables have been calculated in the deformed version of the correspondence. In particular, two, three and $n$-point correlation functions of scalar operators were calculated  in \cite{Fuertes:2009ex} and \cite{Volovich:2009yh}  using the gravity side of the correspondence. It is important to mention that all these operators were dual to point-like strings propagating in the $Sch_5 \times S^5$ background. Extended dyonic giant magnon and spike solutions  were found in
\cite{Georgiou:2017pvi} and their dispersion relations were determined  (in \cite{Zoakos:2020gyb} finite size corrections of those solutions were calculated). The existence of these solutions is in complete agreement with the fact that the theory remains integrable. 
In the same work an exact 
expression, as a function of  the coupling $\lambda$, for the dimensions of the gauge operators dual to the giant magnon solution was conjectured. 
Subsequently, the Penrose limit around a certain null geodesic was utilised in \cite{Georgiou:2019lqh} to study the spectrum of the bosonic strings  in the light-cone 
and show that two of the eigenfrequencies of the bosonic spectrum derived  are in complete agreement with the dispersion relation of the giant magnon solution in the
original Schrodinger background \cite{Georgiou:2017pvi,Georgiou:2018zkt}. 
Inspired by the pp-wave spectrum an improved conjecture about the exact in the t'Hooft coupling dispersion relation
for the magnons in the original  Schr\"{o}dinger background was put forward.
Furthermore, agreement was found between this dispersion relation and the one-loop anomalous dimension of BMN-like operators providing further evidence in favour of the correspondence.
On the field theory side,  the one-loop spectrum of operators belonging in a $SL(2)$ closed sub-sector has been determined in \cite{Guica:2017mtd}. These authors found agreement between the one-loop anomalous dimensions of certain long operators and the string theory prediction (see also \cite{Ouyang:2017yko}).

Subsequently, three-point functions involving two {\it heavy} and one {\it light}  operator were calculated in the Schr\"{o}dinger background by the use of holography  \cite{Georgiou:2018zkt} .
 The  {\it light} operator was chosen to
be the dilaton  while the {\it heavy} states were either the giant magnon or spike solutions constructed in \cite{Georgiou:2017pvi}. These
results are the first in the literature where three-point correlation functions involving
 {\it heavy} states described by extended string solutions were calculated.
The results of \cite{Georgiou:2018zkt} provide the leading term of the correlators in the large $\lambda$ expansion
and are in perfect agreement with the form of the correlator  as this is dictated by non-relativistic conformal invariance.
Additionally, pulsating string solutions in the Schr\"{o}dinger background were constructed in \cite{Dimov:2019koi}.
Finally, Giant Graviton configurations both in the Schr\"{o}dinger background, as well as in its pp-wave limit were found in \cite{Georgiou:2020wwo} and \cite{Georgiou:2020qnh}, respectively. These solutions exhibit an intriguing behaviour, namely  the degeneracy between the point and the Giant Graviton is lifted in favour of the Giant Graviton, as soon as the deformation is turned on. This behaviour is unique since in all the cases studied in the literature the point and the Giant Graviton have always the same energy regardless whether the deformation parameter is zero or non-zero, i.e.   \cite{Pirrone:2006iq,Imeroni:2006rb}.

%%%%%%%%%%%%%%%%%%%%%%%%%%%planofthepaper%%%%%%%%%%%%%%%%%%%%%

In this paper, we continue the study of spacetimes with Schr\"{o}dinger symmetry and their dual dipole-deformed  CFT's. In particular, in section \ref{TheSolution} we derive a new supergravity solution by marginally deforming the Schr\"{o}dinger background presented in appendix \ref{Schrodinger} (see also \cite{Guica:2017mtd}). 
This is accomplished by performing  a series of TsT transformations along two of the five-sphere isometries of the Schr\"{o}dinger solution. The resulting background is integrable for large number of colours $N$ and  depends on two deformation parameters, namely $\mu$ and $\gamma$. In section \ref{FT} we write down the Lagrangian of the field theory dual to our background. It is obtained from the Lagrangian of  $\mathcal {N}=4$ SYM by introducing the appropriate 
star product among the  $\mathcal {N}=4$ SYM fields. The resulting gauge theory has a mild non-locality along one of its light-like directions and is integrable at the planar level. Both of these properties are inherited from the parent Schr\"{o}dinger background.

It is the aim of this work to test  this doubly-deformed new paradigm of the AdS/CFT correspondence and clarify the interplay of the two deformation parameters. In section \ref{Point-like-string} we find a point-like string solution and derive its dispersion relation which depends on both deformation parameters $\mu$ and $\gamma$ in a certain way.
In \ref{coh-states}, we identify the dual field theory operators and in \ref{boundary_conditions} reproduce the leading, in the deformation parameters, terms of the dispersion relation by using the Landau-Lifshitz coherent state approach. This constitutes 
a non-trivial test of the correspondence. 

Subsequently, in section \ref{WL} we evaluate the Wilson loop expectation value, at strong coupling, in the marginally deformed  Schr\"{o}dinger background that we constructed  in \ref{TheSolution}. The Wilson loop consists of two straight lines extending along the time direction and sitting at two points in space separated by a distance ${\mathbb L}$. We calculate the energy of this configuration as a function of the separation length.
This corresponds, as usual, to the potential between two quarks sitting at a distance  ${\mathbb L}$ on the boundary \cite{Maldacena:1998im}.
The energy depends on both deformation parameters and for small values of the separation length ${\mathbb L}$ (with respect to the Schr\"{o}dinger parameter $\mu$) it exhibits confining behaviour, i.e. it is linearly proportional to ${\mathbb L}$ with the constant of proportionality being inversely proportional to $\mu^2$. 
When the separation length is much greater than the Schr\"{o}dinger parameter, the energy is inversely proportional to ${\mathbb L}$ exhibiting the characteristic behaviour of a Wilson loop in conformal theories. We should stress that it is the first time that such a calculation is performed in the 10-dimensional solution of type-IIB supergravity with Schr\"{o}dinger symmetry. In all other cases in the literature the calculation was done in a reduced 5-dimensional background \cite{Armas:2014nea,Araujo:2015qga} and as a result the behaviour of the Wilson loop expectation value was completely different. This is due to the non-trivial winding of the string on the five-sphere which is imposed by consistency of the 10-dimensional equations of motion.

In section \ref{Giant-Graviton} we present the Giant Graviton solution in the background that was constructed in section \ref{TheSolution}. 
The bottomline of the computation is that 
despite the presence of both the Schr\"{o}dinger and the marginal parameters (namely $\mu$ and $\gamma$) in the background, the energy of the Giant Graviton depends only on $\mu$. The parameter $\gamma$ disappears from the computation. As a result, the analysis of the Giant Graviton's energy reduces to the discussion that was presented in \cite{Georgiou:2020wwo}.

In section \ref{pp-wave}, we take the Penrose limit along a certain null geodesic of the Schr\"{o}dinger background.
The geodesic is chosen such that the final geometry depends on both deformation parameters. The main goal of this section is to extract information about the spectrum of our Schr\"{o}dinger background in this particular BMN limit in which string theory is solvable. We fully determine the spectrum of the strings by determining the eigenfrequencies of all 8 transverse physical degrees of freedom of the bosonic string. Their energies give a prediction for the anomalous dimensions of the dual field theory operators as an exact function of the effective coupling $\lambda' =\frac{\lambda}{J^2}$. Four of them depend only on the parameter $\mu$ while the remaining four depend on both deformation parameters $\mu$ and $\gamma$.  Based on the string spectrum on the pp-wave geometry,  we make, in section \ref{spectrum-up}, an educated guess for the {\it exact} in $\lambda$ dispersion relation of the magnon excitations in the original doubly deformed background which provides us with an exact prediction for the dimensions of the dual field theory operators.

In section \ref{concl} we conclude the paper and discuss potential future directions in the framework of the marginally deformed Schr\"{o}dinger holography. The paper is supplemented with two appendixes: In appendix \ref{Schrodinger} we present the Schr\"{o}dinger background before the TsT transformation and in appendix \ref{String_EOM} we write down the conventions for the Polyakov action and the Virasoro constraints.

%%%%%%%%%%%%%%%%%%%%%%%%%%%%%%%%%%%%%%%%%%%%%%%%%%%%%%%%%%%%%%%
\section{Construction of the solution}
\label{TheSolution}

Starting with the type-IIB solution described in section (2.2) of \cite{Guica:2017mtd}, which we review in appendix \ref{Schrodinger}, we perform a TsT transformation along the $U(1)$ isometries generated by the vectors $\partial_\chi$ and $\partial_\phi$. The TsT transformation goes schematically as follows
\begin{itemize}
\item T-duality in the $\chi$-direction: $\chi \rightarrow \tilde{\chi}$
\item Translation in the $\phi$-direction: $\phi \rightarrow \phi + \hat{\g} \, \tilde{\chi}$
\item T-duality in the $\tilde{\chi}$-direction: $\tilde{\chi} \rightarrow \tilde{\tilde{\chi}}$
\end{itemize}
After the third step we rename the coordinate $\tilde{\tilde{\chi}}$ to $\chi$. The deformation parameter $\hat{\g}$ is constant and has units of $[\textrm{length}]^{-2}$. In the following we will express everything in terms of the dimensionless constant $\g = R^2 \hat{\g}$. Let us mention that  
$\hat \gamma$  is related to the field theory deformation 
$\tilde \gamma$ through the following relation
\begin{equation}\label{defrel} 
 2 \, {\hat \gamma}\, = \, \frac{\tilde \gamma}{2 \pi} 
\quad \Rightarrow \quad 
\gamma \, = \, \sqrt{\lambda} \, \frac{\tilde \gamma}{4 \pi } \, , 
\end{equation}
where $\lambda= g^2 N$ is the 't Hooft coupling.

The result of the above process is a solution of the type-IIB supergravity with metric that is given by the line element below

\begin{equation} 
\label{metric}
 \begin{aligned}
  \frac{ds^2}{R^2} & = - \left(  1 + \frac{4 \, \m^2 \, W}{z^4} + \frac{x^2_1 + x^2_2}{z^2} \right) dt^2
  + \frac{1}{z^2} \big(  dx^2_1 + dx^2_2 + dz^2 \big)
  + 4 \, W \, d\chi^2 + d\eta^2
  \\[5pt]
  & + \frac{2}{z^2} \Bigg[  dt \, dv + \g \m W \sin^2\eta \Big(  \cos^2\eta \cos\th d\psi - \big(  1 - \sin^2\eta \cos^2\th \big)d\phi \Big) dt \Bigg]
  \\[5pt]
  & + \frac{W}{4} \sin^2\eta \Big(  4 + \g^2 \sin^2\eta \cos^2\eta \sin^2\th \Big) d\psi^2
  + 4 W \sin^2\eta \, d\chi \Big(  d\psi - \cos\th \, d\phi \Big) 
  \\[5pt]
  & + \frac{\sin^2 \eta}{4} \Big(  d\th^2 + 4 W \, d\phi^2 - 8 W \, \cos\th \, d\phi \, d\psi \Big)
 \end{aligned}
\end{equation}
where $W$ is a function of the angles $\eta$ and $\theta$ given by the expression
\begin{equation} \label{def-W-cal}
 W = \Big[  4 + \g^2 \sin^2\eta \big(  1 - \sin^2\eta \cos^2\th \big)  \Big]^{-1} \, .
\end{equation}
The TsT transformation generates a dilaton that can be expressed in terms of $W$ as follows
\begin{equation} \label{dilaton}
 \Phi = \frac{1}{2} \ln \big(  4 W \big) \, .
\end{equation}
The NS two-form including the new terms arising from the second TsT 
transformation is
\begin{equation} \label{def-B-field}
 \begin{aligned}
  \frac{B_2}{R^2} & = \, \frac{2 \, \m \, W}{z^2} \, dt \wedge 
  \Big(  2 d\chi - \sin^2\eta \cos\th d\phi + \sin^2\eta d\psi \Big)
  \\[5pt]
 & + \, \g \, W \sin^2\eta \, d\phi \wedge 
 \Bigg[  \Big(  1 - \sin^2\eta \cos^2\th \Big) d\chi + 
 \frac{1}{2} \, \sin^2\eta \sin^2\th \, d\psi \Bigg]
 \\[5pt]
 & + \, \g \, W \sin^2\eta \cos^2\eta \cos\th \, d\chi \wedge d\psi \, .
 \end{aligned}
\end{equation}
Finally, for the RR fields we find
\begin{equation}
\label{forms}
 \begin{aligned}
  \frac{F_3}{R^2} & = - \frac{\g}{2} \sin^3\eta \, \cos\eta \, \sin\th \, 
  d\eta \wedge d\th \wedge d\psi
  \\[5pt]
  \frac{F_5}{R^4} & = \frac{4}{z^5} \, dt \wedge dv \wedge dx_1 \wedge dx_2 \wedge dz
  + 2 \, W  \, \sin^3\eta \, \cos\eta \, \sin\th \, d\eta \wedge d\th \wedge d\phi \wedge d\chi \wedge d\psi
  \\[5pt]
  & + \frac{\g \, \m \,W}{z^2} \, \sin^3\eta \, \cos\eta \, \sin\th \, 
  dt \wedge d\eta \wedge d\th \wedge d\psi \wedge 
  \Big(  2 \, d\chi - \sin^2\eta \, \cos\th \, d\phi \Big) \, .
 \end{aligned}
\end{equation}
Notice that when $\g = 0$ one obtains the solution of the appendix \ref{Schrodinger}, while for $\g = \m = 0$ one recovers the original $AdS_5 \times S^5$ solution.

%%%%%%%%%%%%%%%%%%%%%%%%%%%%%%%%%%%%%%%%%%%%%%%%%%%%%%%%%%%%%%%
\section{Dual field theory}
\label{FT}
%%%%%%%%%%%%%%%%%%%%%%%%%%%%%%%%%%%%%%%%%%%%%%%%%%%%%%%%%%%%%%%

In this section, we briefly discuss the field theory which is dual to the type IIB supergravity background that is presented in the previous section. Since this solution is obtained after two sets of TsT transformations\footnote{One should remember that the Scrodinger background of appendix \ref{Schrodinger} is also obtained from the $AdS_5 \times S^5$ solution after a certain TsT transfrormation, i.e. T-duality along the angle $\chi$, shift along a light-like direction and T-duality along the T-dualised isometry $\tilde \chi$. } the dual field theory Lagrangian should be obtained by replacing the ordinary field products in the $\mathcal N=4$ Lagrangian by the following modified star product
\begin{equation}\label{stardefinition}
 \begin{aligned}
 (\Phi _1 \star \Phi _2) (x)=& \left.\,{\rm e}\,^{i \, 
 \frac{\tilde \gamma}{2} \left( Q _{1}^\chi \, Q_2^{\phi} -Q _{1}^{\phi} \,  Q_2^{\chi}\right)} \, 
 {\rm e}\,^{\frac{L}{2} \, \left(\partial_{-1} Q_2^{\chi}- Q_1^{\chi}
 \partial _{-2}\right)} \, \Phi _1(x_1) \, \Phi _2(x_2)\right|_{x_{1,2}=x} 
 \\[3pt]
  = & \, \, {\rm e}\,^{i \, \frac{\tilde \gamma}{2} \, 
  \left(Q _{1}^\chi \, Q_2^{\phi}- Q_{1}^{\phi} \, Q_2^{\chi}\right)} \, 
  \Phi_1 (x+L_2) \, \Phi_2 (x-L_1) \, . 
  \end{aligned}
\end{equation}
One should assign to each field in the theory a dipole length which is proportional to its R-charge along the $\chi$ isometry, namely 
\begin{equation}
L_\Phi \equiv \frac{ L \, Q^\chi_\Phi}{2} \,\hat e_-
\quad {\rm with} \quad 
\hat e_- \, = \, \frac{1}{\sqrt{2}} \, \left(1,-1,0,0\right) \, .
\end{equation}
In the last equation, $\hat e_-$ denotes the vector along the minus null-direction $x^-$ and $Q^\chi_\Phi$ the
R-charge of the field $\Phi$ along the $\chi$ isometry. Notice that the two exponentials in the first line of \eqref{stardefinition} commute since all the participating generators are isometries. This is in complete agreement with the fact that the order of the two sets of TsT transformations is irrelevant, 
i.e. the resulting background is the same regardless which of the two TsT transformations is done first.

As a result, the dipole-deformed SYM Lagrangian becomes
\begin{multline}
  \label{Lagrangian-FT}
  {\cal L}  =  {\rm Tr}\Bigg[- \frac{1}{2} F_{\mu \nu} F^{\mu \nu} +
    2 \, D_{\mu} {\Phi}_{AB} \,  D^{\mu} \Phi^{AB}   + 2 \, i \, \psi^{\alpha}_A \, 
    \sigma_{\alpha \dot{\alpha}}^{\mu} \,  
    (D_{\mu} {\bar{\psi}}^{\dot{\alpha} A}) \, 
    +\\
    2 \, g^2 \left[\Phi^{AB}, \Phi^{CD}\right]_\star \left[{{\Phi}}_{AB}, {{\Phi}}_{CD}\right]_\star -  2 \, \sqrt{2} \, g \, \left([\psi^{\alpha}_A,
      {{\Phi}}^{AB}]_\star
      \psi_{\alpha B} - [{\bar{\psi}}_{\dot{\alpha}}^A, \Phi_{AB}]_\star
      {\bar{\psi}}^{\dot{\alpha}B}\right)\Bigg]
\end{multline}
where the scalar fields with upper indices
$\Phi^{AB}$ ($A,B=1,2,3,4$) are defined as
\begin{equation}
  \label{eq:barPhi}
  {\Phi}^{AB} = \frac{1}{2} \epsilon^{ABCD} \Phi_{CD} \equiv
  {\Phi}^*_{AB},
\end{equation}
and where the covariant derivative  is also defined through the star product
\begin{equation}
 \mathcal{D}_\mu \Phi= \partial_\mu \Phi - i g
[A_\mu,\Phi]_\star 
\end{equation}
with $\Phi$ being any of the fields of the theory.
The theory \eqref{Lagrangian-FT} has a mild non-locality along the null direction $x^-$ (see \eqref{stardefinition}) and non-relativistic conformal invariance along the remaining 3 dimensions.  

The fields $\Phi_{AB}$ are related to the three complex scalars $Z_i,\,\,\,i=1,2,3$ with canonical kinetic term as follows
\begin{equation} \label{comp}
  \Phi_{14} = \frac{Z_1}{2}\, ,\quad
  \Phi_{24} = \frac{Z_2}{2}\, ,\quad
  \Phi_{34} = \frac{Z_3}{2}\, ,\quad
  \Phi_{13} = - \frac{\overline{Z}_2}{2}\, ,\quad
  \Phi_{23} = \frac{\overline{Z}_1}{2} \, ,\quad
  \Phi_{12} = \frac{\overline{Z}_3}{2}.
\end{equation}

We have adopted the ``mostly-minus'' metric $(+,-,-,-)$ and the
following conventions for the gauge group generators
\begin{equation}\label{colcon}
  {\rm Tr}\big(T^a T^b\big) \,=\, \frac{\delta^{ab}}{2}~,~~~
  \left[T^a , T^b \right] \,=\, i f^{abc} T^c~,~~~
  (T^a)^i_j(T^a)^k_l \,=\, 
  \frac{1}{2}\left(
    \delta^i_l \delta^k_j 
    - \frac{1}{N} \delta^i_j \delta^k_l 
  \right).
\end{equation}

In passing, notice that the action \eqref{Lagrangian-FT} is invariant under the following gauge transformations which, however involve the aforementioned star product 
\begin{equation} \label{nlgauge}
 \Phi (x)\rightarrow (U^{-1}\star\Phi \star U)(x)=U^{-1}\left(x+L_\Phi \right)\Phi (x)\, U\left(x-L_\Phi \right)
\end{equation}
instead of the ordinary local gauge transformations. The field with such a transformation law can be thought as a dipole that extends from $x+L_\Phi $ to $x-L_\Phi $ along the minus of the light ray, 
hence the name dipole deformed theory.
Notice also that in order for the star product to be associative, $\Phi_1 \star \Phi_2$ should  be assigned dipole length $L_1 + L_2$, while $L_{\Phi^\dag} = - L_\Phi$ \cite{Dasgupta:2001zu}. %In particular the gauge fields have zero dipole length since they carry zero R-charge under the $\chi$-isometry.  

In more detail, the R-charges of the various fields in the Lagrangian \eqref{Lagrangian-FT} (see also \eqref{comp}) are as follows
\begin{equation}\label{charges1}
 \begin{aligned}
 &Z_1\,,\psi_1: (Q^\chi,Q^\phi,Q^\psi)=(1,1/2,1/2)\\
 &Z_2\,,\psi_2: (Q^\chi,Q^\phi,Q^\psi)=(1,-1/2,1/2)\\
 &Z_3\,,\psi_3: (Q^\chi,Q^\phi,Q^\psi)=(1,0,0)\\
 &A_\mu\,,\psi_4: (Q^\chi,Q^\phi,Q^\psi)=(0,0,0)
  \end{aligned}
\end{equation}
with the conjugate fields $\bar Z_{1}$, $\bar Z_{2}$, $\bar Z_{3}$ and $\bar \psi^A$ having the opposite charges.
These charges can be readily read from the Hopf parametrisation of the undeformed $S^5$ which is as follows
\begin{equation} \label{charges2}
\begin{aligned}
 &Z_1= \sin\eta  \, \sin\frac{\theta }{2} \, e^{i \, \chi} \,  e^{\frac{i}{2} \left(\phi +\psi \right)}, \quad 
 Z_2= \sin\eta \,  \cos\frac{\theta }{2} \, e^{i \, \chi} \,  e^{\frac{i}{2}  \left(-\phi +\psi \right)} 
 \quad \& \quad Z_3=\cos\eta \, e^{i \, \chi} \\
 &ds^2_{S^5}=\sum_{i=1}^3d Z_i^\dagger \,  dZ_i= d\chi^2+ 
 d\eta^2+\sin ^2\eta \,  d\chi \left(d\psi -\cos\theta d\phi \right)\\
 &\qquad \qquad \qquad \qquad \qquad \qquad \qquad +\frac{1}{4} \, \sin ^2\eta  \left(d\theta^2+d\phi^2+d\psi^2-2  \cos \theta \,  d\phi \,  d\psi \right).
  \end{aligned}
\end{equation}
Each of the fermions has the same charge as its supersymmetric bosonic partner. 

%%%%%%%%%%%%%%%%%%%%%%%%%%%%%%%%%%%%%%%%%%%%%%%%%%%%%%%%%%%%%%%
\section{Point-like string and its dispersion relation}
\label{Point-like-string}
\def\cX{\mathcal{X}}
%%%%%%%%%%%%%%%%%%%%%%%%%%%%%%%%%%%%%%%%%%%%%%%%%%%%%%%%%%%%%%%

In the background we described in the previous section, we will now consider point-like string configurations of the 
following form
\begin{equation}\label{ansatze}
 \begin{aligned}
 & \th = \theta_0 \, , \qquad
 \eta = \frac{\pi}{2} \, , \qquad 
 \phi = \psi = x_1 = x_2 = 0 \, ,
\\[5pt]
& 
\chi = \omega \, \tau \, , \qquad 
t = \kappa \, \tau \, , \qquad 
v = \m^2 \, m \, \tau \, , \qquad 
z = z_0 \, .
 \end{aligned}
\end{equation}
Applying the above ansatz to the Plyakov action and from the equations of motion for $z$ and $\theta$ we determine the values of $z_0$ and $\omega$ to be
\begin{equation}
z_0^2 = \frac{\kappa}{m \left(1+\frac{1}{4} \,\gamma^2 \, \sin^2 \theta_0\right)} \quad \& \quad 
\omega = \mu \, m \, \left(1+\frac{1}{4} \, \gamma^2 \, \sin^2 \theta_0\right)
\end{equation}
while $\theta_0$ remains arbitrary. Then from the Virasoro constraint one obtains
\begin{equation} \label{DispersionRelation}
\kappa^2 - \frac{\omega^2}{1+ \frac{1}{4} \, \gamma^2 \, \sin^2 \theta_0}
- \mu^2 \, m^2 
\left(1+\frac{1}{4} \, \gamma^2 \, \sin^2 \theta_0\right) =0 \, .
\end{equation}
We now proceed to rewrite \eqref{DispersionRelation} in terms of the conserved quantities of the solution. 
To this end we calculate the energy, the angular momentum in the five-sphere and the eigenvalue of the mass operator which for our solution read
\begin{equation}\label{charges-sol}
E\, = \, \sqrt{\lambda} \, \kappa \, , \quad 
J\, = \, \sqrt{\lambda} \, 
\frac{\omega}{1 +\frac{1}{4} \, \gamma^2 \, \sin^2 \theta_0} 
\quad  \& \quad 
M \, = \,  \sqrt{\lambda} \, m  
\left( 1 + \frac{1}{4} \, \gamma^2 \, 
\sin^2 \theta_0 \right) \, .
\end{equation}
Substituting now \eqref{charges-sol} in \eqref{DispersionRelation} one obtains
\begin{equation}
\label{DispersionRelation-fin}
E^2 \, = \, \frac{\mu^2 \, M^2}{1 + \frac{1}{4} \, \gamma^2 \, \sin^2 \theta_0} +J^2 \left( 1 + \frac{1}{4} \, \gamma^2 \, 
\sin^2 \theta_0 \right)\, .
\end{equation}
At this point, let us stress that the dispersion relation above gives a prediction at strong coupling for the dimension of the dual field theory operator in the doubly deformed background.
Let us also mention that it depends explicitly on both deformation parameters $\mu$ and $\gamma$.

%%%%%%%%%%%%%%%%%%%%%%%%%%%%%%%%%%%%%%%%%%%%%%%%%%%%%%%%%%%%%%%%%

\subsection{Dispersion relation from the coherent state approach}
\label{coh-states}

In this section, we will show that the dispersion relation \eqref{DispersionRelation-fin}  can be reproduced at leading order in the $\l$ expansion using coherent states. Notice that to this order the dispersion relation
\eqref{DispersionRelation-fin}  becomes
\begin{equation}\label{dr-1loop}
E= J+ J  \, \frac{\gamma^2}{8} \, \sin^2\th_0+ 
\frac{1}{2} \, \frac{\mu^2 \, M^2}{J}
+\mathcal{O}(\mu^2\tilde{\g}^2,\mu^4,\tilde{\g}^4)
\end{equation}
where we have used that  both deformation parameters appearing in the background are proportional to $\sqrt{\l}$. Thus, the sum of the second and the third term in the right hand side of \eqref{dr-1loop} denotes the leading contribution to anomalous dimensions of the dual field theory operators. Notice also that to this order in perturbation theory the two deformation parameters disentangle, i.e. the anomalous dimension is the sum of two terms one involving the  parameter $\mu$ and the other involving $\gamma$. In what follows, we will see that this is in agreement with the expectations coming from the coherent state approach. 
  
The goal of this section is to identify the field theory operators whose dispersion relation is given by \eqref{dr-1loop} and reproduce their anomalous dimension by using the coherent state approach of \cite{Kruczenski:2003gt}. As mentioned in \cite{Guica:2017mtd}, there are two equivalent ways of doing this. The first is to consider spin chains with periodic boundary condition and use the deformed one-loop  Hamiltonian of the theory. The second, which is the one we will be following, is to consider spin chains with doubly twisted boundary conditions but use the undeformed $\mathcal N=4$ one-loop Hamiltonian, instead.

The first step is to identify the operators whose 
dimensions are given by \eqref{DispersionRelation-fin} \& \eqref{dr-1loop}.
By inspecting \eqref{charges1} and \eqref{charges2} one concludes that the operator dual to our point-like string solution should involve the $Z_1$ and $Z_2$ fields but not $Z_3$, namely $J \sin^2\frac{\th_0}{2}$ $Z_1$'s and  
$J \cos^2\frac{\th_0}{2}$ $Z_2$'s 
(see \eqref{charges1} and \eqref{ansatze}). Furthermore, since the string has motion along the $V$-direction the operators should contain the covariant derivative along the minus null-direction $D_-$, as it also happens in the case of the single $\mu$ deformation \cite{Guica:2017mtd}. 
In conclusion, the dual operators will be schematicaly of the form
\begin{equation}\label{operators}
{\cal O}={\rm Tr} (D_-^{S_1} Z_1^{J \sin^2{\th_0 \ov 2}}\, D_-^{S_2} Z_2^{J \cos^2{\th_0 \ov 2}})+...\, ,
\end{equation}
where the dots denote terms in which the derivatives are distributed differently among the scalar fields. The corresponding spin chain will consist of a collection of sites, each of which will accommodate one of the following letters
\begin{equation}\label{letters}
| Z_1\rangle=\Big|{1 \ov 2}\Big \rangle \otimes|\uparrow \rangle,\,\,\,\,\, | Z_2\rangle=\Big|{1 \ov 2} \Big\rangle \otimes| \downarrow \rangle,\,\,\,\,\,
| D^m_-Z_1\rangle=\Big|{1 \ov 2}+m \Big\rangle \otimes|\uparrow \rangle,\,\,\,\,\,| D^m_-Z_2\rangle=\Big|{1 \ov 2}+m \Big \rangle \otimes|\downarrow \rangle\, .
\end{equation}
Thus, we see that at the one-loop level the spin chain is as if it was the product of a $SL(2)\otimes SU(2)$ spin chains. More specifically, the left entry corresponds to the third component of an $SL(2)$ spin at the representation $j={1\ov 2}$ \footnote{The $j={1\ov 2}$ representation  of $SL(2)$ is infinite dimensional with its lowest weight state being $|m_j\, j \rangle =|0\rangle = |{1\ov 2} \, {1\ov 2} \rangle $ while its excited states are $|m_j\, j \rangle = |{1\ov 2} +m\, \,\,{1\ov 2} \rangle ,\,\,\,m>0$.  In \eqref{letters}  the quantum number $j={1\ov 2}$ of the representations have been suppressed for both  $SL(2)$ and $SU(2)$.}  while the right entry corresponds to third component of an $SU(2)$ spin at the representation $j={1\ov 2}$ , respectively. This sector is closed at one-loop but ceases to be so from two-loops on.
The state described by \eqref{operators} is in general a quantum, highly entangled state whose exact form is very complicated. However, when the number of scalars and derivatives is large one may approximate it by low-energy excitations with long wavelengths. The description is then achieved by putting the product of an $SL(2)$ coherent state  times an $SU(2)$ coherent state at each site of the spin chain. The Hamiltonian of the chain is obtained by considering  the continuum limit of the action of the $SL(2)$ and $SU(2)$ one-loop Hamiltonians on the corresponding coherent states.

An alternative, more geometric way of deriving this Hamiltonian is to consider  long classical strings with large charges moving in a $AdS_3\times S^3$ subspace of the complete  $AdS_5\times S^5$ background. For this reason, we choose from \eqref{charges1} a three-dimensinal sphere 
that is parametrised by the angles $\chi$, $\theta$ and $\phi$, setting $\eta=\pi/2$ and $\psi=0$. Then the metric we will use for the $AdS_3\times S^3$ background will be
\begin{equation}\label{repar}
ds^2=-\cosh^2 \frac{\rho}{2} \,dt^2+\frac{1}{4} \, d\rho^2+\sinh^2 \frac{\rho}{2} \,d\phi_1^2+
\frac{1}{4} \left(d\th^2+d\phi^2 \right)+
d\chi^2- \cos \theta \,d\chi \, d\phi.
\end{equation}
The action of the string is given, as usual by the Polyakov action 
\begin{equation}\label{action}
 S={\sqrt{\l}\ov 4 \pi}\int d^2\sigma \,G_{\mu\nu}
 \left(\partial_\tau X^\mu\partial_\tau X^\nu-\partial_\sigma X^\mu\partial_\sigma X^\nu\right)
\end{equation}
and should be supplemented by the Virasoro constraints
\be\label{Vira}
G_{\mu\nu}\partial_\tau X^\mu\partial_\sigma X^\nu=0\quad \& \quad G_{\mu\nu}(\partial_\tau X^\mu\partial_\tau X^\nu+\partial_\sigma X^\mu\partial_\sigma X^\nu)=0.
\ee
The next step is to perform the following change of variables 
\begin{equation} \label{change}
\phi_1 \, = \, \varphi+t \quad \&  \quad 
\chi \, = \, \phi_2-t
\end{equation}
and make the following ansatz for the motion of the string
\begin{equation}\label{ansatz-ft}
t= k\, \tau,\,\,\,\, \varphi=\varphi(\tau,\s),\,\,\,\, \phi_2=\phi_2(\tau,\s),\,\,\,\, \rho=\rho(\tau,\s),\,\,\,\, 
\th=\th(\tau,\s),\,\,\,\,  
\phi= \phi(\tau,\s).
\end{equation}

Subsequently, one should take the Frolov-Tseytlin (F-T) limit which consists in sending $k \rightarrow \infty$ while keeping 
$k \,  \partial_\tau X^\mu={\rm fixed}$, as well as $ \partial_\sigma X^\mu={\rm fixed}$, where $ X^\mu \in \{\rho, \varphi,\phi_2, \th, \phi\}$.
By keeping the leading terms in the F-T limit the first  Virasoro constraint in \eqref{Vira} becomes
\begin{equation}\label{Vira1}
k\, \left(\sinh^2 \frac{\rho}{2} \,\partial_\s\varphi - \partial_\s\phi_2+ \frac{1}{2} \cos \th \, \partial_\s \phi\right) \, = \, 0 \, .
\end{equation}
Similarly, by substituting the Virasoro constraint \eqref{Vira1} in \eqref{action} and keeping again only the leading in the $k$-expansion terms, the action becomes
\begin{equation}\label{action1}
  \begin{aligned}
 S=&{\sqrt{\l}\ov 4 \pi}\int d\tau\int_0^{2 \pi} d\s \,\Bigg[ 2 \, k \, \sinh^2 {\rho \ov 2}\, \partial_\tau \varphi-2 \, k \, \partial_\tau \phi_2 + k \, \cos\th \partial_\tau \phi\\
& \qquad \qquad \qquad \qquad
- \, {1\ov 4} \, \Big[(\partial_\s \rho)^2+ 
\sinh^2 \rho \, (\partial_\s \varphi)^2 + 
(\partial_\s \th)^2 + \sin^2 \th \, 
(\partial_\s \phi)^2\Big]\Bigg]\, .
 \end{aligned}
\end{equation}
From this Lagrangian, one may derive the corresponding Hamiltonian which, after changing the range of the variable $\s$ from $[0,2\pi]$ to $[0,J]$, reads
\begin{equation} \label{Ham}
 \begin{aligned}
 { \cal H}\, =& \, {\lambda \ov 32 \pi^2}\int _0^J d\s\, \Big[-(\partial_\s \vec l)^2+(\partial_\s \vec n)^2\Big]\\[5pt]
 =& \, {\lambda \ov 32 \pi^2}\int _0^J d\s\, \Big[\partial_\s  l_+ \, \partial_\s  l_- 
 - (\partial_\s  l_0)^2+\partial_\s  n_+ \, \partial_\s  n_-+(\partial_\s  n_0)^2\Big]
\end{aligned}
\end{equation}
where $\vec l$ and $\vec n$ parametrise a point on a 2-dimensional hyperboloid and on a two dimensional sphere, respectively. More specifically
\begin{equation}\label{par1}
 \begin{aligned}
&\vec l=(l_x,l_y,l_0)=(\sinh\rho \cos\varphi,\sinh\rho \sin\varphi,\cosh\rho) \quad {\rm with} \quad \vec l\cdot \vec l=1=l_0^2-l_x^2-l_y^2\\[5pt]
&\vec n=(n_x,n_y,n_0)=(\sin\th \cos\phi,\sin\th \sin\phi,\cos\th) \quad {\rm with} \quad
\vec n\cdot \vec n=n_0^2+n_x^2+n_y^2=1 
  \end{aligned}
\end{equation}
with $l_\pm = l_x \pm i \, l_y$. 
Two important comments are in order. The first is that the Hamiltonian \eqref{Ham} is precisely the Hamiltonian one gets by taking the continuum limit of the one-loop Hamiltonians of the $SL(2)$ and $SU(2)$ closed subsectors of ${\cal N}=4$ SYM theory. Secondly, notice that the Hamiltonian governing the dynamics of the spin chain \eqref{letters} at one-loop order is the sum of two independent terms. This is in agreement with the fact that at the leading order the two deformation parameters $\mu$ and $\g$ disentangle in the string dispersion relation \eqref{dr-1loop}.

The Poisson brackets of the $l_i$ and $n_i$ are determined from the commutation relations of the $SL(2)$ and $SU(2)$ algebras respectively, after the usual replacement $\{\, ,\}_{P.B.} \to -i  [\,,]$. 
 For the  $SL(2)$ case and by taking into that $S^i\rightarrow n_i/2$ \cite{Guica:2017mtd} , 
 we find that 
\begin{eqnarray}\label{PoissonSL}
 \Big\{l_0(\sigma ),\, l_\pm(\sigma ')\Big\}&=&
 \mp \, 2 \, i \, l_\pm(\sigma ) \, 
 \delta \left(\sigma -\sigma '\right)
\nonumber \\[3pt]
\Big\{l_+(\sigma ),\, l_-(\sigma ')\Big\}&=& 
4 \, i \, l_0(\sigma ) \, 
\delta \left(\sigma -\sigma '\right) \, .
\end{eqnarray}
The Hamilton equations for (\ref{Ham}) by the use of  (\ref{PoissonSL}) are the Landau-Lifshitz equations for the classical  (non-compact) ferromagnet
\begin{eqnarray}\label{eom1}
\partial _tl_0 =\{l_0,\cal H\}& = & - \, \frac{i \, \lambda }{16 \, \pi ^2} \, \left(l_- \, \partial _\sigma ^2l_+
 - l_+ \, \partial _\sigma ^2l_-\right)
\nonumber \\[3pt]
\partial _t l_\pm =\{l_\pm,\cal H\}& = & \mp \, \frac{i \, \lambda }{8 \, \pi ^2} \, \left(l_0 \, \partial _\sigma^2 l_\pm - l_\pm \, \partial _\sigma ^2l_0\right)
\end{eqnarray}
It is an easy task to check that time evolution preserves the constraint $\vec l\cdot \vec l=1$ and thus only two of the three equations for the components of $\vec{l}$ are independent.
Similarly, for the $SU(2)$ case one gets
\begin{eqnarray}\label{PoissonSU}
 \Big\{n_0(\sigma ), \, n_\pm(\sigma ')\Big\} & = & 
 \mp \, 2 \, i \, n_\pm(\sigma ) \, \delta \left(\sigma -\sigma '\right)
\nonumber \\[3pt]
\Big\{n_+(\sigma ), \, n_-(\sigma ') \Big\} &=& - \, 4 \, i \, n_0(\sigma ) \, \delta \left(\sigma -\sigma '\right).
\end{eqnarray}
The Hamilton equations for (\ref{Ham}) by the use of  (\ref{PoissonSU}) are the Landau-Lifshitz equations for the classical  spin-1/2 ferromagnet
\begin{eqnarray}\label{eom2}
 \partial _tn_0=\{n_0,\cal H\}&=&- \, \frac{i \, \lambda }{16 \, \pi ^2} \, \left(n_- \, \partial _\sigma ^2n_+
 -n_+ \, \partial _\sigma ^2n_-\right)
\nonumber \\[3pt]
\partial _tn_\pm =\{n_\pm,\cal H\}& = & \pm \, \frac{i \, \lambda }{8 \, \pi ^2} \, \left(n_0 \, \partial _\sigma ^2n_\pm- n_\pm \, \partial _\sigma ^2n_0\right)
\end{eqnarray}
It is again straightforward to verify that time evolution preserves the constraint $\vec n\cdot \vec n=1$. 

The last step is to solve \eqref{eom1} and  \eqref{eom2} with the appropriate  twisted boundary conditions. These solutions will be presented in the next section.

%%%%%%%%%%%%%%%%%%%%%%%%%%%%%%%%%%%%%%%%%%%%%%%%%%%%%%%

\subsection{Doubly twisted boundary conditions}
\label{boundary_conditions}

The last step before deriving the dispersion relation \eqref{dr-1loop} from the coherent state approach is to set up the appropriate twisted boundary conditions and subsequently seek solutions of  \eqref{eom1} and  \eqref{eom2} given these boundary conditions. Given the form of the one-loop states \eqref{letters} the operator implementing the twisted boundary conditions is the product of two operators, one for the $SL(2)$ part and one for the $SU(2)$ part,
\begin{equation}\label{bc}
 \left|s_{J+1}\right \rangle\otimes\left|\hat s_{J+1}\right \rangle= S\left|s_{1}\right \rangle\otimes\hat S \left|\hat s_{1}\right \rangle,
  %\quad {\rm with} \quad{\cal S}=S\otimes \hat S, 
\end{equation}
where $S$ corresponds to the $SL(2)$ part while $\hat S$ corresponds to the $SU(2)$ part.

For the equations \eqref{eom1} we closely follow \cite{Guica:2017mtd}. There are only minor differences which originate from the definitions of the coherent states. While in \cite{Guica:2017mtd} one has that $\langle \vec l |S^i|\vec l \rangle=-{1\ov 2}l_i$ we follow the normalisations of \cite{Georgiou:2011qk} in which $\langle \vec l |S^i|\vec l \rangle={1\ov 2}l_i$. 
In this case the twisted boundary conditions for the spin chain read  \cite{Guica:2017mtd}
\begin{equation}\label{bc-1}
 \left|s_{J+1}\right\rangle=S\left|s_1\right\rangle
 \quad {\rm with} \quad
 S=\,{\rm e}\,^{iL\left(
  P_{-\,1}{\mathbf Q^\chi}-Q^\chi_{1}\mathbf{P}_-
 \right)},
\end{equation}
with the boldface letters denoting the total charges of the spin chain as a whole. 
This equation implies the following twisted boundary conditions for the spin operators of the $SL(2)$ sub-chain. Recall that the total $R$-charge along the $\chi$ isometry of the spin chain is equal to half of its length, i. e.  $J$ and $P_-$ is identified with $-iS^-$
\begin{equation}
 S^i_{J+1}\, = \,{\rm e}\,^{- \, L \, J \, S^-_1} \, S_1^i\,{\rm e}\,^{L \, J \, S^-_1} \, .
\end{equation}
By employing the anti-commutators for the  spin operators one derives for the last similarity transformation the following relations
\begin{equation}
S^-_{J+1} \, = \,  S_1^- \, , \quad 
S^0_{J+1} \, = \, S_1^0-L \, J\,S_1^- \quad \& \quad 
S^+_{J+1}\, =\, S_1^+ - 2 \, L \, J\,S_1^0+ L^2 \, J^2\,S_1^-
\end{equation}
which translate to the following quasi-periodic boundary conditions
\begin{eqnarray}
 l_-(t,\sigma +J)&=&l_-(t,\sigma )
\nonumber \\
l_0(t,\sigma +J)&=&l_0(t,\sigma )-L \, J\,l_-(t,\sigma )
\nonumber \\
l_+(t,\sigma +J)&=&l_+(t,\sigma )-2 \, L \, J\,l_0(t,\sigma )+L^2 \, J^2\,l_-(t,\sigma ).
\end{eqnarray}
At this point, let us mention that solutions satisfying these twisted boundary conditions are necessarily complex.
Then by making the ansatz  \cite{Guica:2017mtd}
\begin{equation}\label{ansatz}
l_0 \, = \, \alpha +\beta \, \sigma  \quad \& \quad 
l_- \, = \, \frac{2 \, i \, M}{J}
\end{equation}
one can immediately see that the equation for $l_-$ is satisfied while $l_+$ is uniquely determined through the constraint $l_0^2-l_+ l_-=1$.
Notice also that $\alpha $ and $\beta $ may depend on time.
Using the aforementioned constraint the equation for $l_0$ (first equation in \eqref{eom1}) becomes
\begin{equation}
 \partial _tl_0 \, = \, - \, \frac{i \, \lambda }{16 \, 
 \pi ^2}\,\partial _\sigma ^2l_0^2 
\quad \Rightarrow \quad
\alpha \, = \, - \, \frac{i \, \lambda }{8 \, \pi ^2}\,\beta ^2 \, t \, . 
\end{equation}
The boundary conditions \eqref{bc-1} are then enough to fix $\beta $
\begin{equation}
\beta \, = \, - \, \frac{2 \, i \, L \, M}{J} \, .
\end{equation}
Plugging now the solution in the first two terms of \eqref{Ham}, we obtain the part of the one-loop anomalous dimension that depends on the deformation parameter $\mu$
\begin{equation}\label{anom-1}
 \Delta_1 \, = \, - \, \frac{\lambda }{32 \, \pi ^2}\,J \, \beta ^2 \, = \, \frac{\lambda \, L^2 \, M^2}{8 \, \pi ^2 \, J} \, = \, {1\ov 2} \, {\mu^2 \, M^2 \ov J}
\end{equation}
which agrees perfectly with the third term in the string prediction \eqref{dr-1loop}. To derive the last equality we have taken into account that $\mu=\sqrt{\lambda} \, {L \ov 2 \, \pi}$.

We now turn to the solution of the second set of equations \eqref{eom2}. 
In this case the twisted boundary conditions for the spin chain read  
\begin{equation}\label{bc-2}
 \left|\hat s_{J+1}\right\rangle=\hat S\left|\hat s_1\right\rangle
 \quad {\rm with} \quad 
 \hat S=\,{\rm e}\,^{i \, \tilde \gamma \, \left(
  Q_1^\chi{\mathbf Q}^\phi - Q^\phi_{1}{\mathbf Q}^\chi
 \right)}
\end{equation}
with the boldface letters denoting the total charges of the spin chain as a whole.
This equation implies the following twisted boundary conditions for the spin operators of the $SL(2)$ sub-chain. Recall that the total $R$-charge along the $\chi$ isometry of the spin chain is equal to half of its length, i. e.  $J$, as well as that $Q^\phi_{1}$ corresponds to $\hat S_1^0$ (see \eqref{charges1})
\begin{equation}
 \hat S^i_{J+1} \, =\,{\rm e}\,^{i \, \tilde \gamma \, J \,  \hat S^0_1} \, \hat S_1^i\,{\rm e}\,^{- \, i \, \tilde \gamma \, J \, \hat S^0_1} \, .
\end{equation}
By employing the anti-commutators for the  spin operators for the $SU(2)$ algebra now, one derives for the last similarity transformation the following relations
\begin{equation}
\hat S^0_{J+1} \, = \, \hat S_1^0 \, , \quad 
\hat S^+_{J+1} \, = \,  \hat S_1^+{\rm e}\,^{- \, 2 \,  i \, \tilde \gamma \, J \, \hat S^0_1} \quad  \& \quad 
\hat S^-_{J+1} \, = \, {\rm e}\,^{ 2 \, i \, \tilde \gamma \, J  \,  \hat S^0_1} \, \hat S_1^- 
\end{equation}
which translate to the following quasi-periodic boundary conditions
\begin{eqnarray}\label{twisted-bc2}
n_0(t,\sigma +J)&=&n_0(t,\sigma )\nonumber \\[5pt]
n_\pm(t,\sigma +J)&=&{\rm e}\,^{\mp \,  i \, \tilde \gamma \, J \,  n_0(t,\s)} \, n_\pm(t,\sigma ) \, .
\end{eqnarray}
 We now make the following ansatz 
\begin{equation}\label{ansatz2}
n_0 \, = \, \cos\frac{\theta_0}{2} \quad \& \quad
n_\pm \, = \, \sin\frac{\theta_0}{2} {\rm e}\,^{\pm \, i \, \left(a_1 \, t +\b_1 \,\s \right)} 
\quad {\rm with} \quad  n_0^2+n_+n_-=1 \, . 
\end{equation}
One can immediately check that the first equation in \eqref{eom2} is satisfied.
The equation for $n_+$ in \eqref{eom2} determines $\a_1$ in terms of $\b_1$, namely
\begin{equation}
\a_1 \, = \, - \, {\lambda \,\b_1^2 \ov 8 \, \pi^2} \, \cos{\th_0 \ov2} \, .
\end{equation}
Finally, the second equation in the twisted boundary conditions \eqref{twisted-bc2} specify $\b_1$ to be
\begin{equation}
\b_1 \, = \, - \, \tilde \gamma \, \cos{\th_0 \ov2} \, .
\end{equation}
Plugging this value in the last two terms of \eqref{Ham} one gets
for the part of the one-loop anomalous dimension of our operators which depends on the deformation parameter $\gamma$
\begin{equation}\label{anom-2}
 \Delta_2 \, = \, \frac{\lambda }{32 \, \pi ^2}\,J \, 
 \sin^2{\th_0 \ov 2}\, \beta_1 ^2 \, = \, 
 \frac{J\, \lambda  }{32 \, \pi ^2 }{\tilde \gamma^2 \ov 4}\sin^2\th_0 \, = \, J \,  {\g^2 \ov 8} \, \sin^2\th_0
\end{equation}
which agrees perfectly with the third term in the string prediction \eqref{dr-1loop}. To derive the last equality we have taken into account \eqref{defrel}.

Summing up \eqref{anom-1} and \eqref{anom-2} we get the complete one-loop anomalous dimension of our operators $\D^{(1)}=\D_1+\D_2$ which as already mentioned is in perfect agreement with the string prediction \eqref{dr-1loop}.

%%%%%%%%%%%%%%%%%%%%%%%%%%%%%%%%%%%%%%%%%%%%%%%%%%%%%%%%

\section{Wilson loop calculation}
\label{WL}

In this section we will evaluate the expectation value of a certain Wilson loop operator (WL) in the marginally deformed  Schr\"{o}dinger background 
that we constructed  in section \ref{TheSolution}. 
The Wilson loop consists of two straight lines extending along the time direction and sitting at two points in space separated by a distance $\mathbb{L}$. We calculate the energy of this configuration as a function of the separation length.
This corresponds, as usual, to the potential between two quarks sitting at a distance  $\mathbb{L}$ on the boundary \cite{Maldacena:1998im}. The standard prescription for the holographic computation
that was introduced in \cite{Maldacena:1998im} dictates the minimization of the Nambu-Goto action for a fundamental 
string propagating into the dual supergravity background. The string endpoints are lying on the two sides of the 
Wilson loop.  The presence of the Schr\"{o}dinger deformation parameter imposes the presence of a 
relative angle between the quarks. 
Contrary to other TsT deformed geometries that are known in the literature, there is no consistent ansatz 
for the trajectory of the Wilson loop for which all the angles of the deformed five-sphere are set to constant values. 

We will perform the computation in Poincare coordinates. For this reason we quote explicitly the form of the background metric (we have set the radius of the spacetime $R=1$)
\begin{equation} 
\label{metric-Poincare}
 \begin{aligned}
  ds^2 & = -  \frac{4 \, \m^2 \, W}{z^4}  dx_+^2
  + \frac{1}{z^2} \big(  dx^2_1 + dx^2_2 + dz^2 \big)
  + 4 \, W \, d\chi^2 + d\eta^2
  \\[5pt]
  & + \frac{2}{z^2} \Bigg[  dx_+ \, dx_- + \g \m W \sin^2\eta \Big(  \cos^2\eta \cos\th d\psi - \big(  1 - \sin^2\eta \cos^2\th \big)d\phi \Big) dx_+ \Bigg]
  \\[5pt]
  & + \frac{W}{4} \sin^2\eta \Big(  4 + \g^2 \sin^2\eta \cos^2\eta \sin^2\th \Big) d\psi^2
  + 4 W \sin^2\eta \, d\chi \Big(  d\psi - \cos\th \, d\phi \Big) 
  \\[5pt]
  & + \frac{\sin^2 \eta}{4} \Big(  d\th^2 + 4 W \, d\phi^2 - 8 W \, \cos\th \, d\phi \, d\psi \Big)
 \end{aligned}
\end{equation}
where the definition for $W$ is in \eqref{def-W} and the expression for the dilaton is the same as in \eqref{dilaton}.
For the WL computation we also need the expression of the NS two-form
\begin{equation} \label{B-field-Poincare}
 \begin{aligned}
  B_2 & = \, \frac{2 \, \m \, W}{z^2} \, dx_+ \wedge 
  \Big(  2 d\chi - \sin^2\eta \cos\th d\phi + \sin^2\eta d\psi \Big)
  \\[5pt]
 & + \, \g \, W \sin^2\eta \, d\phi \wedge 
 \Bigg[  \Big(  1 - \sin^2\eta \cos^2\th \Big) d\chi + 
 \frac{1}{2} \, \sin^2\eta \sin^2\th \, d\psi \Bigg]
 \\[5pt]
 & + \, \g \, W \sin^2\eta \cos^2\eta \cos\th \, d\chi \wedge d\psi \, .
 \end{aligned}
\end{equation}
In order to make contact with the other Wilson loop computations that appear in the literature, we perform the following change of variables to the metric \eqref{metric-Poincare} and the NS two-form \eqref{B-field-Poincare}
\begin{equation} \label{WL-change-variables}
x_{-} = \frac{1}{\sqrt{2}} \left(- t +  \zeta_1 \right) \, , \quad 
x_{+} = \frac{1}{\sqrt{2}} \left(t + \zeta_1 \right) \quad \& \quad 
z = \frac{1}{u} 
\end{equation}
and now the boundary is located at $u\rightarrow \infty$. We have checked that the following ansatz is consistent 
with the corresponding equations of motion
\begin{equation}
\eta = \frac{\pi}{2} \, , \quad 
\theta = \frac{\pi}{2} \, , \quad 
\phi =  \psi = x_2 = 0 \quad \& \quad
\zeta_1 = 0 \, . 
\end{equation}
Setting these values to the metric \eqref{metric}, after the change of variables that we introduced in 
\eqref{WL-change-variables}, we arrive to the following reduced four-dimensional metric
\begin{equation} \label{reduced-metric}
ds^2 = - u^2 \left[ 1+\frac{\tilde{\mu}^2 \,u^2}{1+\tilde{\gamma}^2}\right] dt^2 + u^2 dx_1^2 + \frac{du^2}{u^2} +
\frac{d\chi^2}{1+\tilde{\gamma}^2}
\quad {\rm with} \quad 
\tilde{\gamma}  = \frac{\gamma}{2}  \quad \& \quad  \tilde{\mu}  = \frac{\mu}{\sqrt{2}} \, .
\end{equation}
while the NS-two form of equation \eqref{B-field-Poincare} becomes
\begin{equation}
B_2 = \frac{ \tilde{\mu} \, u^2}{1+ \tilde{\gamma} ^2} \, dt \wedge d\chi \, .
\end{equation}
Notice that in this section $\tilde{\gamma}$ is not the field theory deformation parameter appearing in equation \eqref{defrel}. 
The Nambu-Goto action for the fundamental string in the presence of a B-field has the following form
\begin{equation}
S = \frac{1}{2\, \pi} \int d\tau \, d\sigma \left[\sqrt{-\det \gamma_{\alpha  \beta}} - 
\frac{1}{2} \epsilon^{\alpha  \beta} \, B_{\alpha  \beta}\right]
\end{equation}
with the expressions for the symmetric and the antisymmetric contributions are given as usual be the 
following expressions
\begin{equation}
\gamma_{\alpha  \beta} = g_{MN} \partial_\alpha x^M \partial_\beta x^N \quad \& \quad
B_{\alpha  \beta} = B_{MN} \partial_\alpha x^M \partial_\beta x^N \, . 
\end{equation}
Notice that in the trajectories we are considering there is a contribution from the antisymmetric part, contrary to the 
considerations of \cite{Akhavan:2008ep, Armas:2014nea,Araujo:2015qga} 
where the backgrounds that are probed are not embedded in string theory, that is they are 5-dimensional. 
The embedding ansatz for the string trajectory is the following
\begin{equation}
x_1= \sigma = x\, , \quad u = u(x) \quad \& \quad \chi = \chi(x) \, . 
\end{equation}
Applying this ansatz to the reduced metric \eqref{reduced-metric}, the Nambu-Goto action for the string becomes
\begin{equation}
S = \frac{1}{2\, \pi} \int dx \left[\sqrt{\big. g(u) (\partial_x u)^2 + f(u) + h(u)  (\partial_x \chi)^2} -  
w(\sigma) \, \partial_x \chi \right]
\end{equation}
where the different functions are defined as follows
\begin{equation}
g(u) = 1 + \frac{\tilde{\mu}^2 \,u^2}{1+\tilde{\gamma}^2} \, , \quad f(u) = u^4 g(u) \, , \quad  
h(u) = \frac{u^2}{1+\tilde{\gamma}^2} \, g(u)  \quad \& \quad w(u) = \frac{\tilde{\mu} \, u^2}{1+\tilde{\gamma}^2} \, . 
\end{equation}
Conservation of energy and angular momentum lead to the following first order differential equations
\begin{equation} \label{energy-momentum-conservation}
\frac{f}{\sqrt{\big. g \, u'^2 + f + h \, \chi'^2}} = \alpha_1 
\quad \& \quad 
\frac{h \, \chi'}{\sqrt{\big. g \, u'^2 + f + h \, \chi'^2}} -w = \kappa
\end{equation}
where $\kappa$ and $\alpha_1$ are constants.
The string develops a minimum at the turning point in which the derivative of $u$ with respect to $x$ vanishes. Renaming the
constants in such a way that this point is denoted as $u_0$, amounts to the following relation between 
$\alpha_1$ and $u_0$
\begin{equation}
\alpha_1 = u_0^2 \, \sqrt{\big.1- 2 \, \kappa \, \tilde{\mu}} \, \sqrt{1- \xi^2} \quad {\rm with} \quad \xi = \frac{\kappa}{u_0} \sqrt{\frac{1+\tilde{\gamma}^2}{1- 2 \, \kappa \, \tilde{\mu}}} \, . 
\end{equation}
Notice that there is no way for the other constant $\kappa$ to be fixed. Solving the equations in 
\eqref{energy-momentum-conservation} for $x$ and integrating from the turning point $u_0$ until the boundary, 
we obtain the expression for the length of the Wilson loop. The string, that penetrates in the bulk, connects a quark 
with an antiquark on the boundary, placed at the positions $x=\mathbb{L}/2$ and $x=-\mathbb{L}/2$ respectively. 
The expression for the length reads 
\begin{eqnarray} \label{WL-length}
\mathbb{L} & = &  2 \, u_0^2 \, \sqrt{1- \xi^2} \int_{u_0}^{\infty} \frac{du}{u^2 \sqrt{\big. \left(u^2 -u_0^2\right) 
\left[u^2 +u_0^2 \left(1-\xi^2\right)\right]}}
\nonumber \\ 
&= & \frac{2}{u_0 \sqrt{\left(1- \xi^2\right) \left(2- \xi^2\right)}} \left[(2- \xi^2) \, {\bf E}(k)-  {\bf K}(k)\right] 
\quad {\rm with} \quad k = \frac{1- \xi^2}{2- \xi^2}
\end{eqnarray}
where $ {\bf K}(k)$ and $ {\bf E}(k)$ are the complete elliptic integrals of the first and second kind, respectively.
The calculation of the angle breaks in two parts: There is a finite contribution that vanishes if we set the constant 
$\kappa$ to zero (recall that $\kappa$  is related to the conservation of the ``angular momentum") 
and an infinite contribution (i.e. divergent integral) that needs to be regularized in the same fashion as the energy. 
More specifically we have 
\begin{equation}  \label{WL-angle-full}
\Delta \chi  = \Delta \chi^{\rm finite} + \Delta \chi^{\rm reg} 
\end{equation}
and the finite part is given by the following integral that can be calculated analytically in terms of the elliptic integral
$ {\bf K}(k)$
\begin{eqnarray}
\Delta \chi^{\rm finite}  &= &  \frac{2}{\sqrt{\big. 1- 2 \, \kappa \, \tilde{\mu}}}
\int_{u_0}^{\infty} \frac{\kappa \left(1+\tilde{\gamma}^2\right)}{\sqrt{\big. \left(u^2 -u_0^2\right) 
\left[u^2 +u_0^2 \left(1-\xi^2\right)\right]}} \, du
\nonumber \\ 
&= & 2 \, \sqrt{1+ \tilde{\gamma}^2} \, \frac{\xi}{\sqrt{2 - \xi^2}} \,  {\bf K}(k) \, . 
\end{eqnarray}
To regularize the divergent integral in the infinite contribution, we subtract the angle of two straight strings that in the 
Schr\"{o}dinger part of the geometry extend from the boundary until the distance $\zeta$. The result of the 
calculation is
\begin{eqnarray} \label{WL-angle-infinite}
\Delta \chi^{\rm reg}  & =  & \frac{2\, \tilde{\mu}}{\sqrt{\big. 1- 2 \, \kappa \, \tilde{\mu}}}
\Bigg[\int_{u_0}^{\infty} \frac{ u^2 \, du}{\sqrt{\big. \left(u^2 -u_0^2\right) 
\left[u^2 +u_0^2 \left(1-\xi^2\right)\right]}}  - \int_{\zeta}^{\infty} \frac{ u \, du}{\sqrt{\big. u^2 - \zeta^2}} \Bigg]
\nonumber \\ 
&= & - \frac{2 \, \tilde{\mu} \, u_0}{\sqrt{\big. 1- 2 \, \kappa \, \tilde{\mu}}}
 \frac{1}{\sqrt{2- \xi^2}} \left[(2- \xi^2) \, {\bf E}(k)-  {\bf K}(k)\right] \quad {\rm with} \quad
 \zeta = u_0 \, \xi \, .
\end{eqnarray}
The regularization term in \eqref{WL-angle-infinite} subtracts the infinity of the first integral close to the boundary 
but does not add any finite piece. 
Notice that the regularized contribution to the angle in  \eqref{WL-angle-infinite} is directly proportional to the 
Schr\"{o}dinger parameter $\tilde \mu$ and vanishes if we set $\tilde \mu$ to zero.
The total energy of the Wilson loop is also divergent and has to be regularized. The result of the computation is
\begin{eqnarray} \label{WL-energy}
E_{q \bar{q}} & =&  \frac{1}{\pi} \frac{1- \kappa \, \tilde{\mu}}{\sqrt{\big. 1- 2 \, \kappa \, \tilde{\mu}}} 
\Bigg[\int_{u_0}^{\infty} \frac{ u^2 \, du}{\sqrt{\big. \left(u^2 -u_0^2\right) 
\left[u^2 +u_0^2 \left(1-\xi^2\right)\right]}}  - \int_{\zeta}^{\infty} \frac{ u \, du}{\sqrt{\big. u^2 - \zeta^2}} \Bigg]
\nonumber \\ 
&= & - \frac{u_0}{\pi \, \sqrt{2- \xi^2}} \frac{1- \kappa \, \tilde{\mu}}{\sqrt{\big. 1- 2 \, \kappa \, \tilde{\mu}}} 
\left[(2- \xi^2) \, {\bf E}(k)-  {\bf K}(k)\right] \, . 
\end{eqnarray}
Setting the Schr\"{o}dinger parameter $\tilde \mu$ to zero and imposing that 
$\kappa = \ell \, u_0 \left(1+{\tilde \gamma}^2\right)^{-1/2}$ where $\ell$ is a constant with 
$0 <\ell <1 $, we end-up with the expressions for the 
length, angle and energy that were presented in section 5.2 of \cite{Hernandez:2005xd} 
(see also \cite{Hernandez:2005zx}).

\begin{figure}[ht!]
\centering
\includegraphics[width=0.45\textwidth,height=6cm]{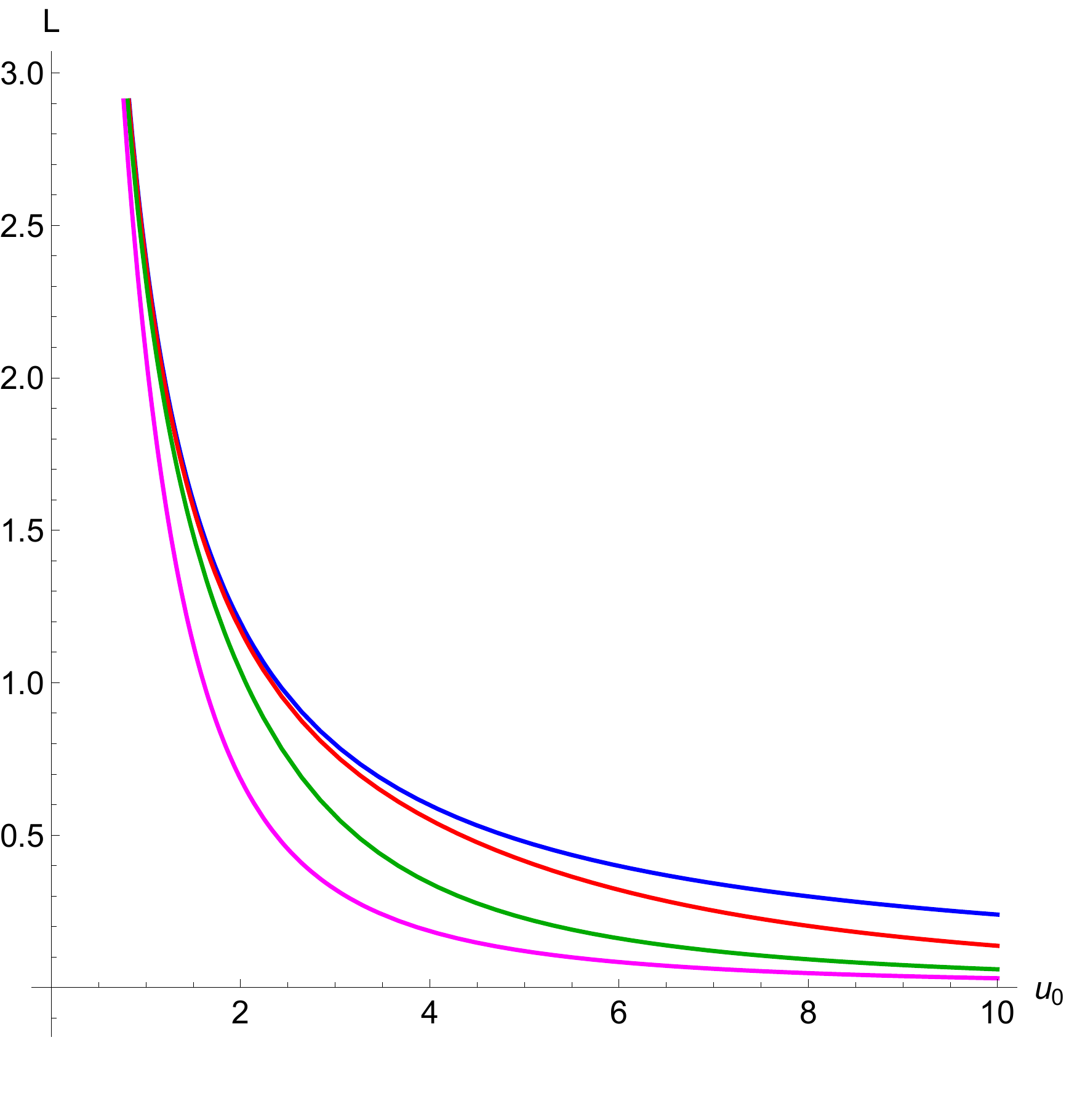}
\hspace{1cm}
\includegraphics[width=0.45\textwidth,height=6cm]{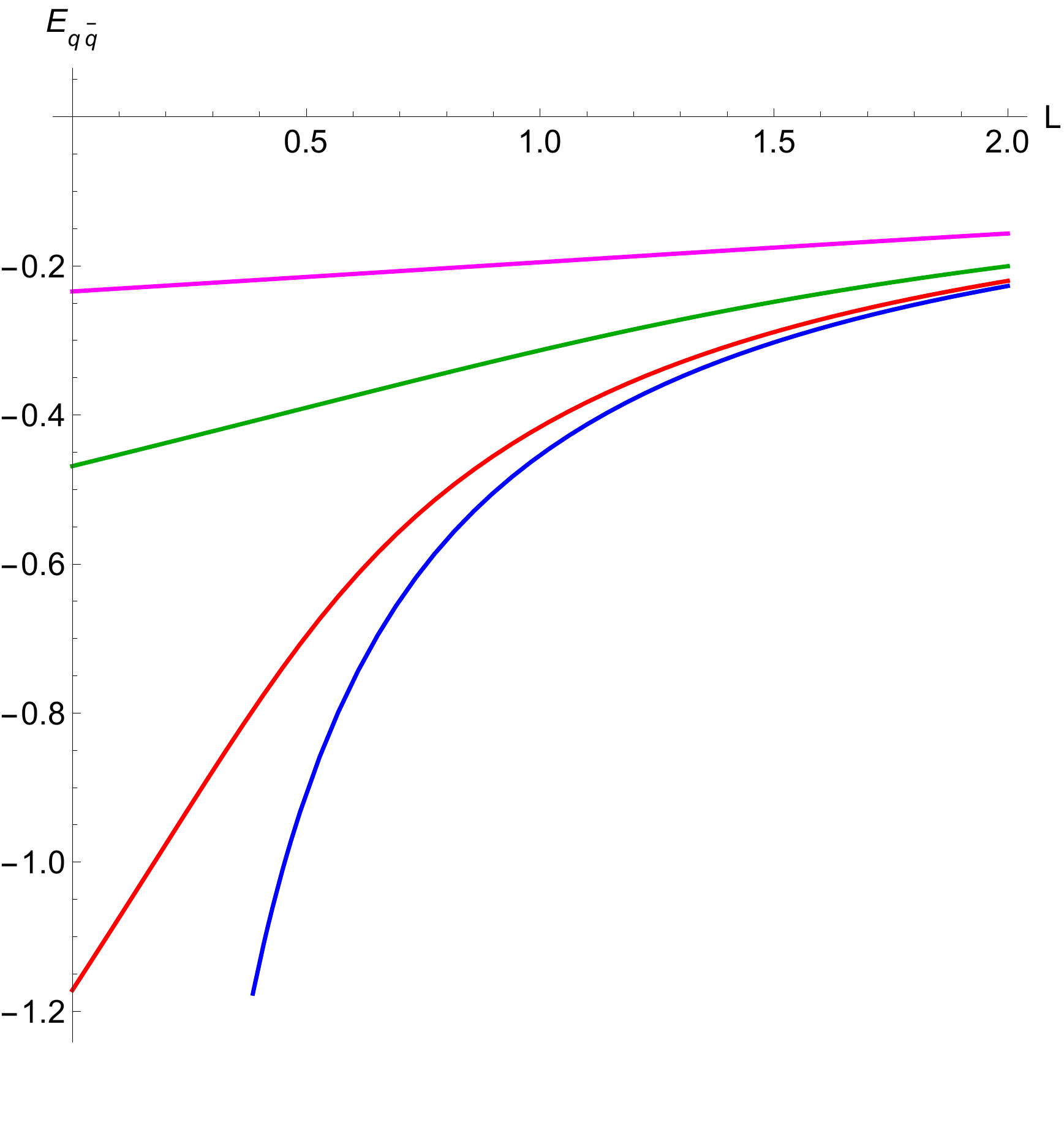}
\includegraphics[width=0.45\textwidth,height=6cm]{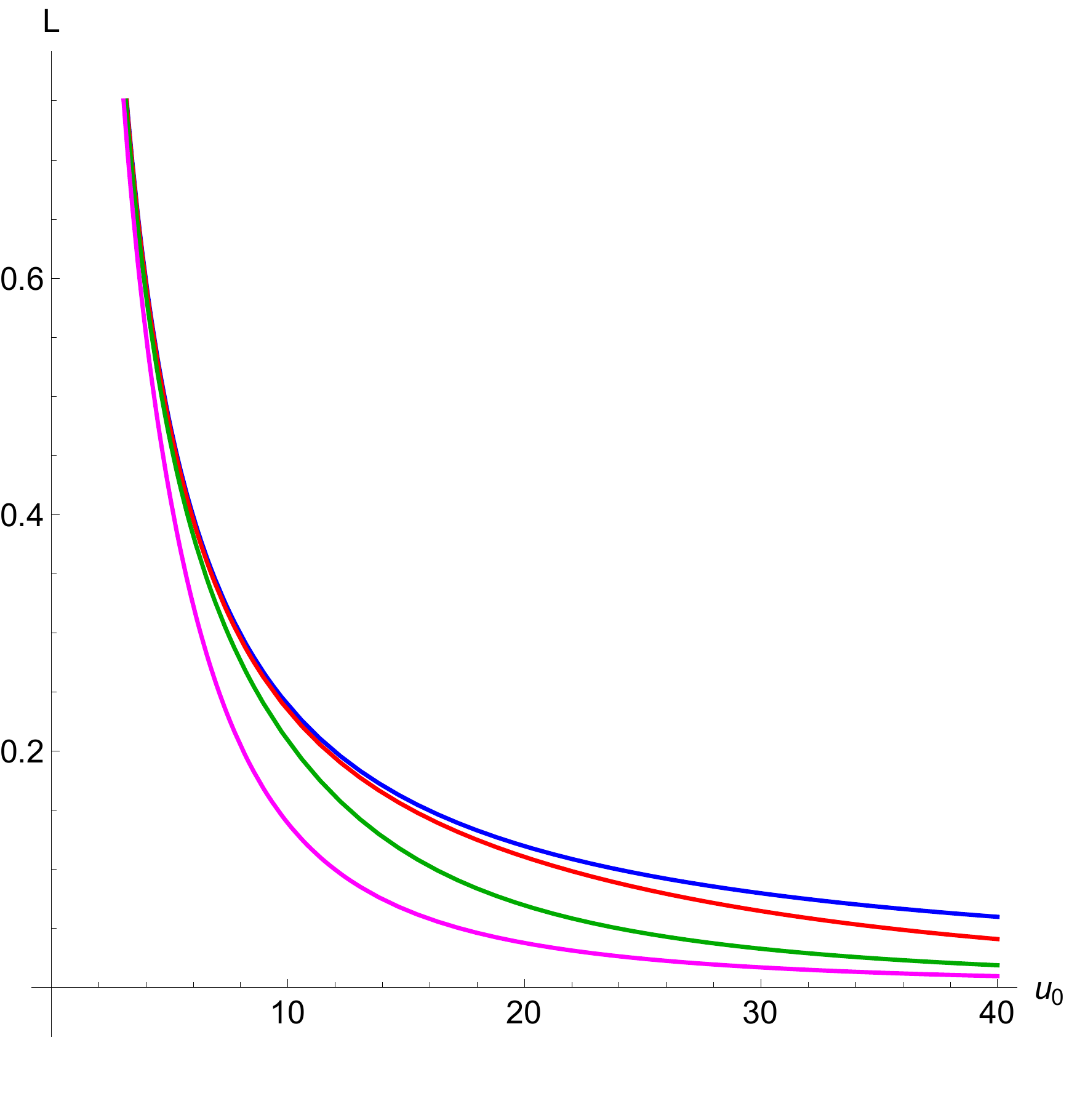}
\hspace{1cm}
\includegraphics[width=0.45\textwidth,height=6cm]{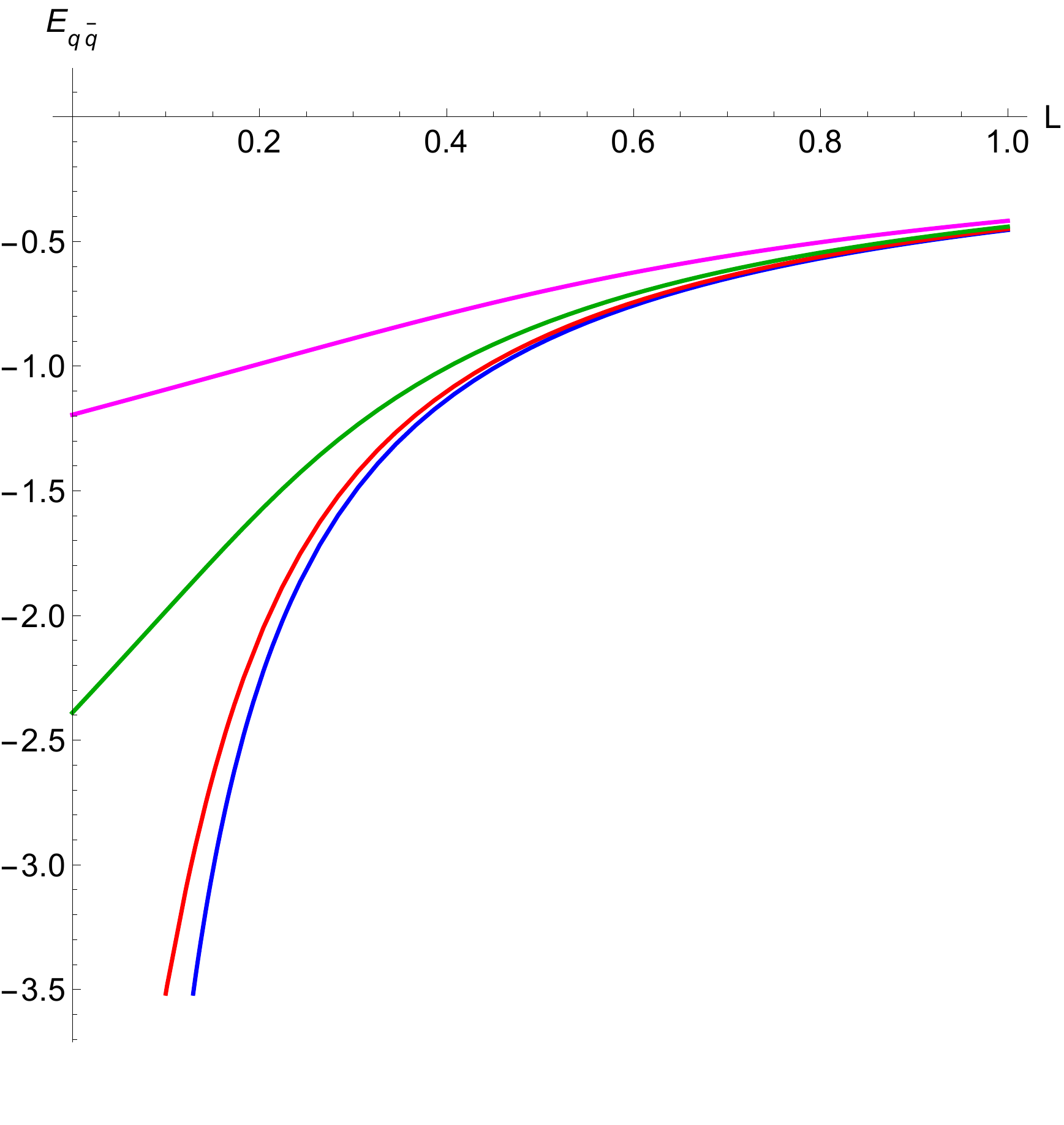}
\caption{Plots of the length of the WL as a function of the penetration parameter $u_0$ (left panels) and of the energy of the WL as a function of the length (right panels) for two different values of the 
deformation parameter $\tilde \gamma$. More specifically, the upper two plots are for 
$\tilde \gamma=0$ and the lower two plots are for $\tilde \gamma=5$. The angle between the quarks along the five sphere is kept fixed at the value $\Delta \chi = {\pi \over 16} \sqrt{1+ \tilde{\gamma}^2}$ . 
Different colors correspond to different values of the deformation parameter $\tilde{\mu}$: blue for $\tilde{\mu}=0$, red for $\tilde{\mu}=0.2$, green for $\tilde{\mu}=0.5$ and magenta for $\tilde{\mu}=1$}
 \label{WL1}
\end{figure}

To proceed with the computation, we have to solve for the auxiliary parameters $\kappa$ and $u_0$ in terms of the 
boundary distance $\mathbb L$ and the angle $\Delta \chi $ and then substitute this solution/identification to the 
expression of the energy in \eqref{WL-energy}. 
This can be done analytically only for large values of $u_0$, since in this case the expressions for the elliptic integrals can be expanded. 
For generic values of $u_0$ we do the following: We fix the 
angle $\Delta \chi$ to a specific value and by solving numerically equation \eqref{WL-angle-full} we determine 
all the pairs $(\kappa, u_0)$ that correspond to this specific value.  Afterwards we substitute those values 
to the expressions for the Wilson loop length and energy, namely  \eqref{WL-length} and  \eqref{WL-energy}.
In figure  \ref{WL1} we have plotted the length as a function of the parameter $u_0$ (left panels) and the energy as a 
function of the length (right panels) for two different values of the 
deformation parameter $\tilde \gamma$. More specifically, the upper two plots are for $\tilde \gamma=0$ and the lower two plots are 
for $\tilde \gamma=5$. The angle is fixed to the value  
$\Delta \chi = {\pi \over 16} \sqrt{1+ \tilde{\gamma}^2}$.   
The different colors in the curves correspond 
to different values of the deformation parameter $\tilde{\mu}$. Plots with similar behavior can be obtained by changing the 
value of the angle and of $\tilde \gamma$.

Fitting the data of the plots from figure \ref{WL1} in the region where $u_0$ is small, we see
that for both values of the 
deformation parameter $\tilde{\gamma}$ and for all the values of the 
deformation parameter $\tilde{\mu}$, the length is inverse 
proportional to $u_0$ and the energy is inverse proportional to the length. Based on that observation we conclude 
that for large values of $\mathbb L$, when the Wilson loop is probing the far interior of the background and the quarks are far 
away from each other, the geometry has $AdS$ characteristics. 

\begin{figure}[ht!]
\centering
\includegraphics[width=0.45\textwidth,height=6cm]{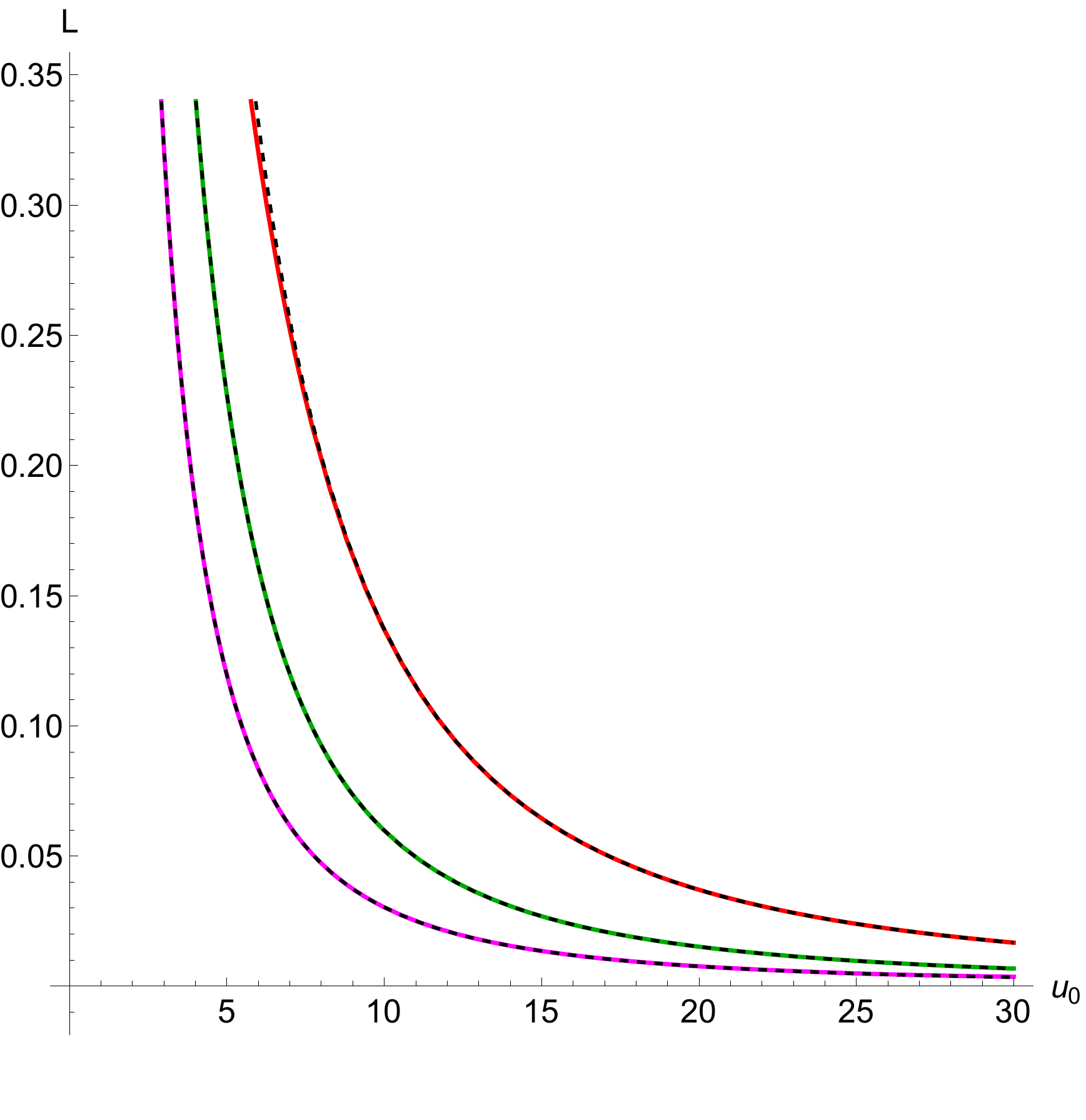}
\hspace{1cm}
\includegraphics[width=0.45\textwidth,height=6cm]{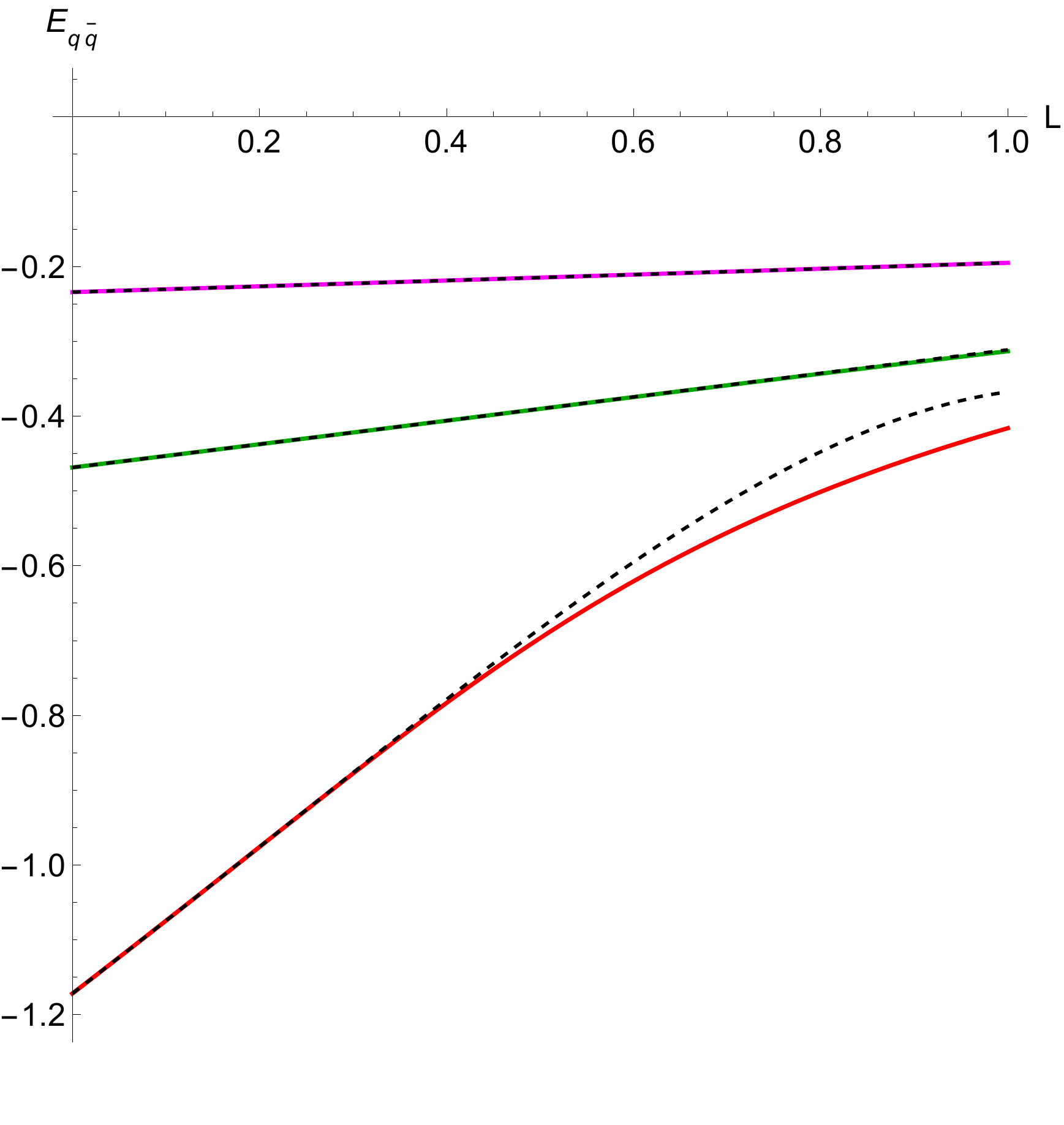}
\includegraphics[width=0.45\textwidth,height=6cm]{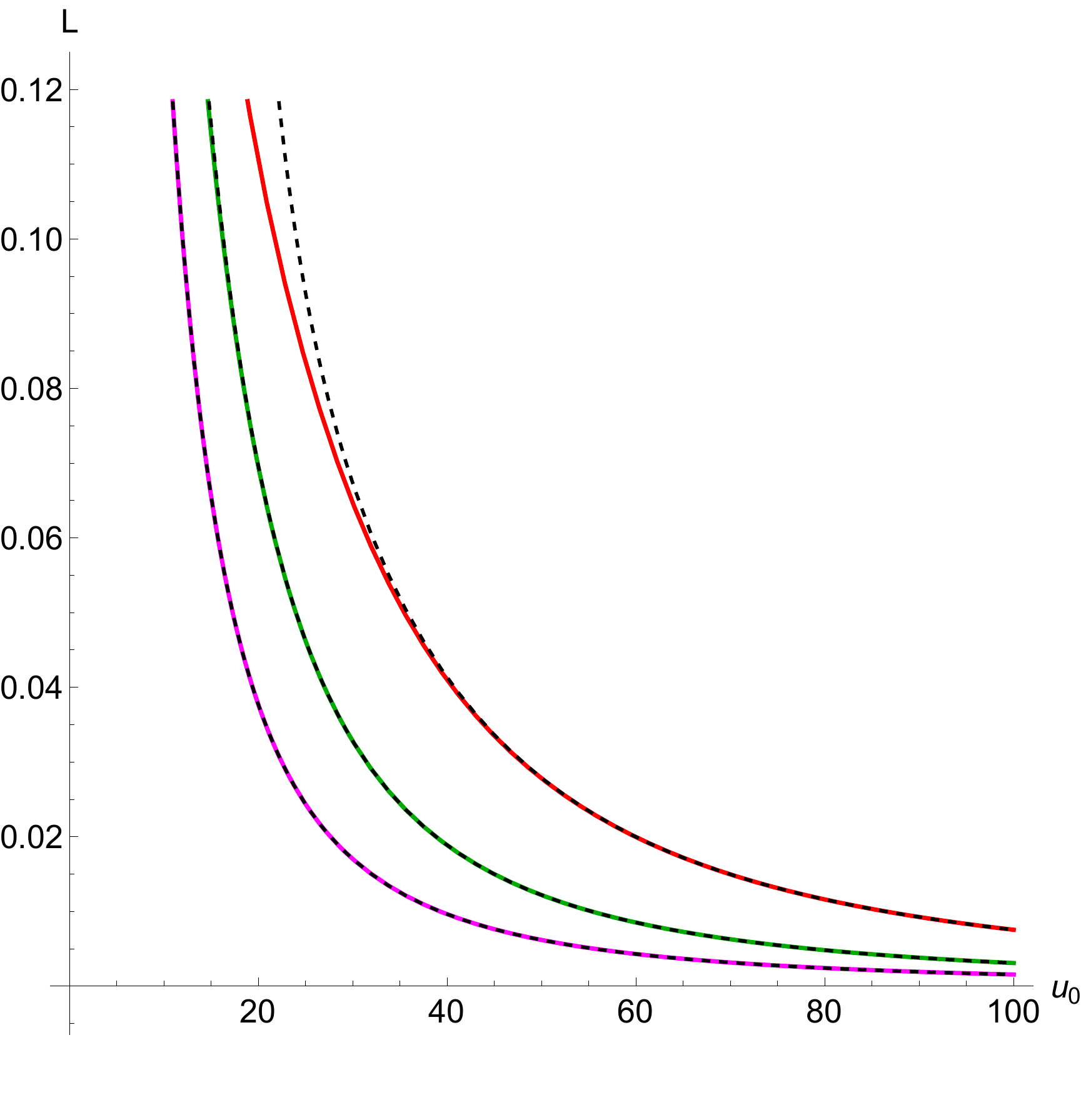}
\hspace{1cm}
\includegraphics[width=0.45\textwidth,height=6cm]{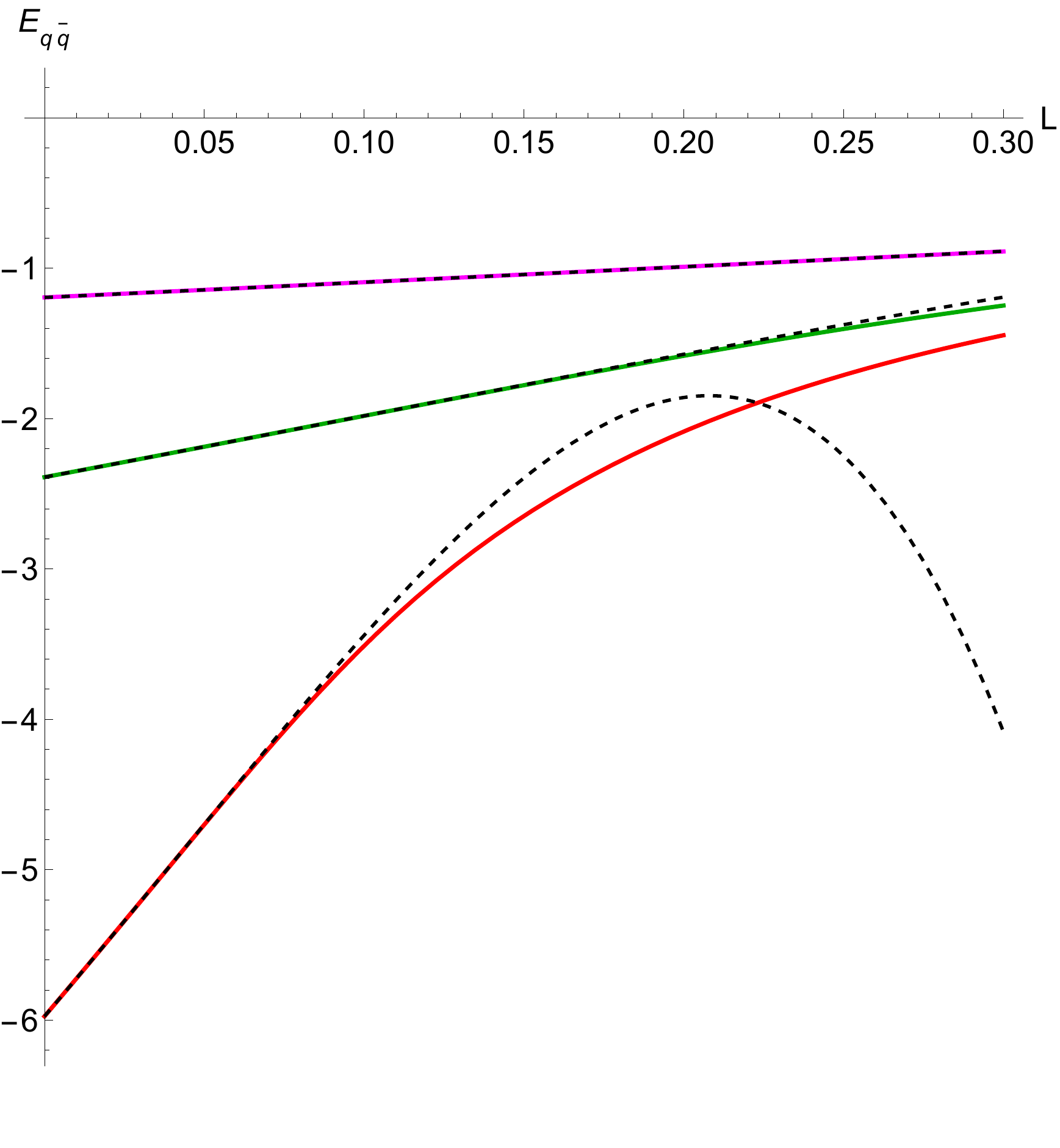}
\caption{Zooming in of the plots of figure \ref{WL1} in the region of large values of $u_0$ or equivalently of small
values of $\mathbb{L}$. The solid lines are the result of the numerical evaluation, while for the dotted black lines we have used the 
perturbative expressions, namely equations  \eqref{WL-length-expansion} and  \eqref{WL-energy-length-expansion}.}
 \label{WL2}
\end{figure}

This picture changes considerably when the value of $u_0$ increases. This can be seen explicitly, since the elliptic integrals in equations \eqref{WL-length}, \eqref{WL-angle-full} and \eqref{WL-energy} can be expanded perturbatively. 
To simplify the analytic expressions, the constant value of the angle is denoted as 
\begin{equation}
\Delta \chi = \ell \, \pi \, \sqrt{1+ \tilde{\gamma}^2} 
\end{equation}
where $\ell$ is a constant. Inspired by the fitting of the numerical data that we have already obtained, 
we introduce the following ansatz for $\kappa$ in  \eqref{WL-angle-full} 
\begin{equation} \label{kappa-expansion}
2 \, \kappa \, \tilde{\mu} = 1 - \frac{1+ \tilde{\gamma}^2}{4} \, \frac{1}{\tilde{\mu}^2 u_0^2 }
- \frac{\left(1+ \tilde{\gamma}^2 \right)^2}{8} \, \frac{1 -2 \, \ell}{\tilde{\mu}^4 u_0^4 }
- \frac{\left(1+ \tilde{\gamma}^2 \right)^3}{64} \,  \frac{3 -28 \, \ell+30 \,  \ell^2}{\tilde{\mu}^6 u_0^6 }
\end{equation}
and expanding in large values of $u_0$ we determine the coefficients. Substituting \eqref{kappa-expansion} in 
\eqref{WL-length} and \eqref{WL-energy} we obtain the following expressions for the length
\begin{equation} \label{WL-length-expansion}
\mathbb{L} = \frac{\pi}{\tilde{\mu} \, u_0^2} \sqrt{\big.\left(1+ \tilde{\gamma}^2 \right)(1-\ell)} \Bigg[ 1 - 
\frac{1+ \tilde{\gamma}^2}{16} \, \frac{\ell +7}{\tilde{\mu}^2 u_0^2 } 
+ \frac{ \left(1+ \tilde{\gamma}^2 \right)^2}{512} \, \frac{99 -26 \, \ell+135 \,  \ell^2}{\tilde{\mu}^4 u_0^4 }\Bigg]
\end{equation}
and for the energy
\begin{equation}  \label{WL-energy-expansion}
E_{q\bar q } = - \frac{1- \ell}{4 \, \tilde{\mu}} \sqrt{1+ \tilde{\gamma}^2} \Bigg[1-\frac{1+\tilde{\gamma }^2}{2\, \tilde{\mu }^2 \, u_0^2}
+\frac{\left(1+\tilde{\gamma }^2\right)^2}{64} \, \frac{11+ 17 \ell}{ \tilde{\mu }^4 \, u_0^4} \Bigg]
\end{equation}
as expansions in inverse powers of $u_0$. Combining  \eqref{WL-length-expansion} and \eqref{WL-energy-expansion}
we obtain the following expression for the energy in powers of the length
\begin{equation}  \label{WL-energy-length-expansion}
E_{q \bar q} = - \frac{1- \ell}{4 \, \tilde{\mu}} \sqrt{1+ \tilde{\gamma}^2} + 
\frac{(1+ \tilde{\gamma}^2)\sqrt{1-\ell}}{8 \, \pi \, \tilde{\mu}^2} \, \mathbb{L}+
\frac{3 \left(1+ \tilde{\gamma}^2\right)^{3\over 2} \, \left(1-5 \, \ell \right)}{256 \, \pi^2 \, \tilde{\mu}^3} \, \mathbb{L}^2 \cdots \, . 
\end{equation}
In figure \ref{WL2} we compare the perturbative expansions of the length and of the energy with respect to the numerical 
counterparts. Keeping the same number of terms in the expansion, the overlap of the two curves (numerical and perturbative) is maintained for larger values of the length, the lower the value of the deformation parameter 
$\tilde \mu$ is. 

\begin{figure}[ht!]
\centering
\includegraphics[width=0.45\textwidth,height=6cm]{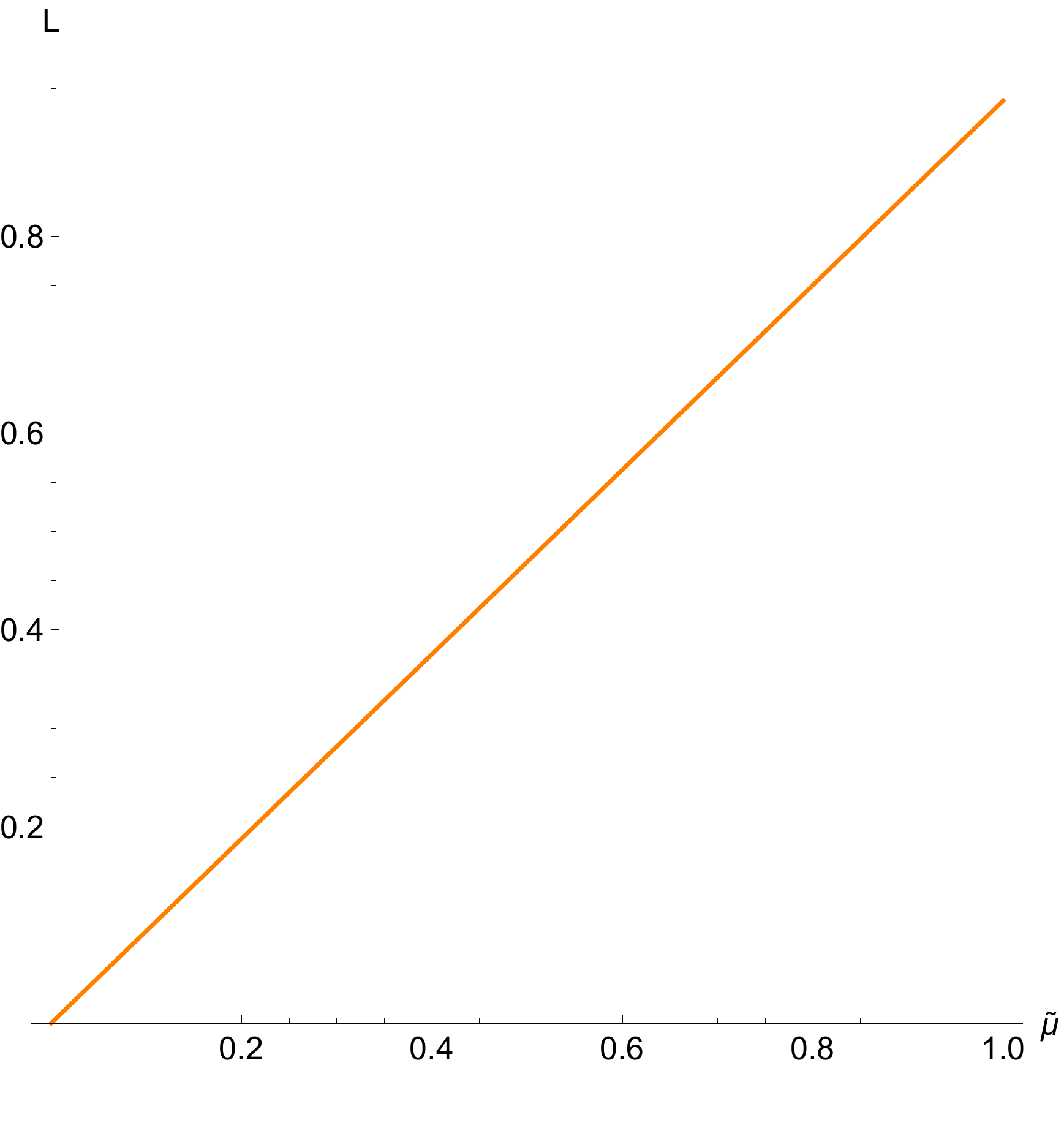}
\hspace{1cm}
\includegraphics[width=0.45\textwidth,height=6cm]{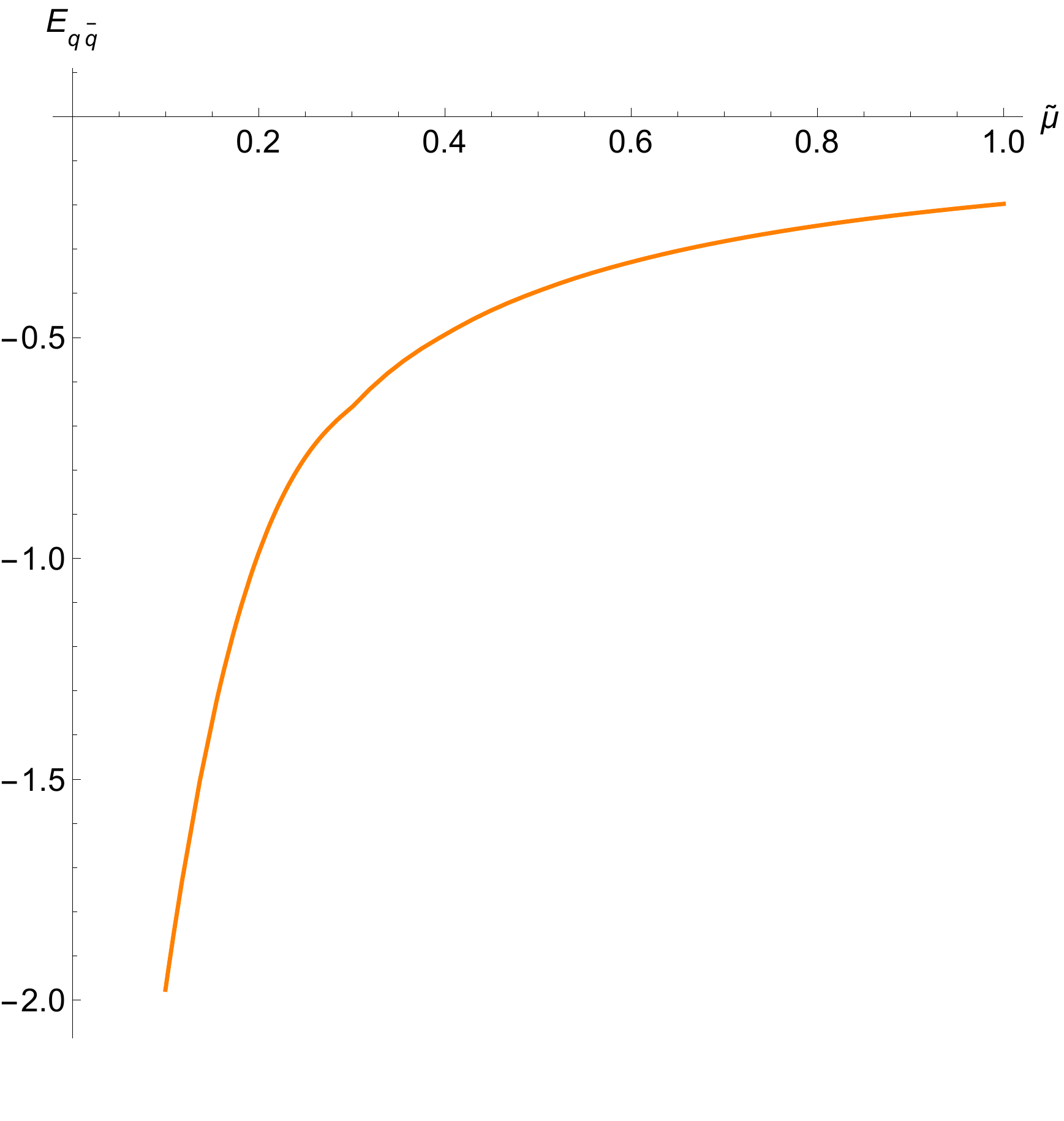}
\caption{Plots of the length (left panel) and of the energy (right panel) of the inflection point 
as functions of the deformation parameter $\tilde{\mu}$. When the length is smaller, the Wilson loop has a 
confining behavior. Fitting the data we confirm that the length is linear in $\tilde{\mu}$ while the energy is inverse proportional.}
 \label{WL3}
\end{figure}

There are various interesting conclusions to draw from the preceding analysis and especially from 
equation \eqref{WL-energy-length-expansion}. As immediate observation is that as the length of the Wilson loop 
decreases, the energy instead of decreasing to minus infinity which would be the expected $AdS$ behavior, 
flows to a finite value that is inversely proportional to the deformation parameter $\tilde{\mu}$. 
As a result, by increasing the value of $\tilde{\mu}$ we can bring the quarks 
close to each other  with less energy cost and in the limit of $\tilde{\mu} \rightarrow \infty$ we have complete screening.
Another striking feature of  \eqref{WL-energy-length-expansion} is that for small values of the boundary distance 
the energy is linear in the length and a confining behavior arises. This behavior remains dominant for the Wilson loop 
until an inflection point in the plot of the energy versus the length is reached. 
This is not easily seen by looking in the plots of figure \ref{WL1} and one has to draw the first and the second derivative 
of the energy with respect to the length.  It is after that point that 
the characteristic $AdS$ structure that we described previously emerges. In figure \ref{WL3} we have plotted the length and the energy of the inflection point as functions of the deformation parameter $\tilde{\mu}$. The length 
is almost linear in $\tilde{\mu}$ while the energy is inverse proportional. Notice that the inflection point we calculate 
numerically is different from the inflection point that one could calculate using the perturbative expansion 
for the energy in  \eqref{WL-energy-length-expansion}. This is not surprising since as can be seen from figure \ref{WL3},
for large enough values of $\tilde{\mu}$ the inflection point is not anymore close to the $\mathbb{L} \rightarrow 0 $ limit.

The analysis of the behavior of the Wilson loop is telling us that in order to probe the effect of the deformation 
parameter $\tilde{\mu}$ in the geometry, we have to stay close to the boundary (where $u_0$ is large and 
consequently $\mathbb{L}$ is small) and not penetrate deep in the IR. This is explained by the fact that in the reduced metric \eqref{reduced-metric} the term 
that contains the effect of the  $\tilde{\mu}$ becomes important when $u$ is large, i.e. close to the boundary. 
This is a behavior that distinguishes the Schr\"{o}dinger solution from all the other TsT transformed backgrounds.

%%%%%%%%%%%%%%%%%%%%%%%%%%%%%%%%%%%%%%%%%%%%%%%%%%%%%%%%%%%%%%%%%%%%%%%%%%%%%%%%%%%%%%%%%%%%%%%%%

\section{Giant Graviton}
\label{Giant-Graviton}

In this section we will present the Giant Graviton solution in the background that was constructed in section \ref{TheSolution}. 
The bottom-line of the computation is that 
despite the presence of both the Schr\"{o}dinger and the marginal parameters (namely $\mu$ and $\gamma$) in the background, the energy of the Giant Graviton depends only on $\mu$. The parameter $\gamma$ disappears from the 
computation. As a result, the analysis of the Giant Graviton's energy reduces to the discussion that was presented in \cite{Georgiou:2020wwo}. 
Here we will examine the technicalities behind the $\gamma$ cancellation. 
Notice the following: in the literature it is known that in the marginally deformed backgrounds 
the energy of the Giant Graviton
is {\it blind} to the presence of the $\gamma$ deformation (see for example \cite{Pirrone:2006iq,Pirrone:2008av,Imeroni:2006rb}). However, since in the background of section  \ref{TheSolution}
there is a mixing between the two deformations, a term with a $\gamma$ dependence could potentially {\it survive}. In the 
following we will see why this is not the case.  

To proceed with the computation along the lines of \cite{Grisaru:2000zn}, we perform the following change of variables 
in \eqref{metric} 
\begin{equation}
\rho = \sin \eta \, . 
\end{equation}
To construct the WZ term of the D3-brane action that describes the Giant Graviton, we need to determine the 
RR potentials. For $F_3$ in \eqref{forms} we determine $C_2$ as follows  
\begin{equation}
F_3 = dC_2 \quad \Rightarrow \quad 
\frac{C_2}{R^2} \, = \, -\,  \frac{1}{8} \, \gamma \, \rho^4 \, \sin \theta \, d\rho \wedge d\theta \wedge d \phi
\end{equation}
while from the definition for $F_5$ 
 \begin{equation}
F_5  = dC_4 - H_3 \wedge C_2 
\end{equation}
and \eqref{forms}, we obtain an expression for $C_4$ that has a smooth zero deformation limit
\begin{equation} \label{C4_potential}
\frac{C_4}{R^4}  = - \, \frac{1}{z^4} \, dt \wedge dv \wedge dx_1 \wedge dx_2 + \frac{\rho^4}{8} \sin \theta \, d\theta \wedge d \phi
\wedge d\chi \wedge d \psi + B_2 \wedge C_2 \, . 
\end{equation}

Now that all the potentials are known, we have to consider the action of a probe D3-brane to describe Giant 
Graviton solutions in \eqref{metric}. The action has the usual form 
\begin{equation} \label{D3-action}
S_{{\rm D}3} = - T_3 \int d^4 \xi  \, e^{-\Phi} \sqrt{\big. - \det {\cal P}\Big[g-B+2 \pi \alpha' F \Big] } +
 T_3 \int \sum_q {\cal P}\Big[ C_q \wedge e^{- B+2 \pi \alpha' F} \Big]
\end{equation}
where ${\cal P}$ denotes the pullback on the brane worldvolume directions and $T_3$ is the tension of the D3-brane.
An important comment is in order: Notice that expanding the exponential of the WZ term, the combination of the potentials that will contribute to the action is $C_4 - B_2 \wedge C_2$. From \eqref{C4_potential} we observe that this 
combination is deformation independent. This means that the WZ contribution to the action will be identical to the 
undeformed case and all the potential deformation dependence will come only from the DBI term.  To see whether this 
happens, we have to consider a consistent ansatz for the D3-brane embedding. 
The D3-brane probe we consider extends along the following directions 
\begin{equation}
\xi_0 = \tau \, ,\quad 
\xi_1 = \theta \, ,\quad 
\xi_2 = \phi \quad \& \quad 
\xi_3 = \psi
\end{equation}
while the ansatz for the embedding is the following
\begin{equation} \label{D3-ansatz-gg}
t =  \, \frac{\tau}{R} \, ,\quad 
v = \nu  \, \tau \, ,\quad 
\vec{x} = 0  \, ,\quad
z = z_0 \, ,\quad
\rho = \frac{\rho_0}{R}
\quad \& \quad
\chi= \omega \, \tau
\end{equation}
with $\rho_0$ being the size of the graviton in the internal space.

Substituting the ansatz \eqref{D3-ansatz-gg} in the D3-brane action \eqref{D3-action} and integrating, an impressive 
cancellation occurs and, we obtain the on-shell action and the Lagrangian of \cite{Georgiou:2020wwo} 
that only depend on the Schr\"{o}dinger deformation parameter $\mu$. The presence of the dilaton in the string frame DBI term is essential for the $\gamma$ cancelation in the action. The Lagrangian becomes
\begin{equation} \label{Lagrangian}
L_{{\rm D}3} = -\, \frac{N}{R^4} \, 
\Bigg[ \rho_0^3 \, \sqrt{1 - \Gamma - \left(R^2 - \rho_0^2\right) \, 
\left(\omega^2 - \frac{\Delta^2}{R^2}\right)} - \rho_0^4 \, \omega \Bigg]
\end{equation}
where the different constants are defined as follows
\begin{equation}
\kappa = \frac{1}{R} \ , \quad 
\Gamma =  \frac{2 \, R}{Z_0^2} \, \nu \ , \quad 
\Delta = \frac{\mu}{Z_0^2}
\quad \& \quad
2 \, \pi^2 \, T_3 = \frac{N}{R^4} \, . 
\end{equation}
From this point on the analysis is identical to the case that only the Schr\"{o}dinger deformation is present and can be found 
in \cite{Georgiou:2020wwo}. Notice that if one performs a stability analysis around the Giant Graviton solution in 
\eqref{D3-ansatz-gg}, the vibration modes will depend on both deformation parameters $\mu$ and $\gamma$. 
Support on this argument is coming from the study of the spectra of frequencies  (for both the scalar and the 
gauge fields) in the $\gamma$ deformed backgrounds \cite{Pirrone:2006iq,Pirrone:2008av,Avramis:2007wb}.

%%%%%%%%%%%%%%%%%%%%%%%%%%%%%%%%%%%%%%%%%%%%%%%%%%%%%%%%%%%%%%%%%%%%%%
\section{PP-wave background and the string spectrum}
\label{pp-wave}

In this section we consider the Penrose limit of the deformed solution that was constructed in section \ref{TheSolution}, around a geodesic that sits at $\eta = 0$. Notice that this geodesic and the corresponding dispersion relation do not depend on the deformation parameter $\gamma$. In order to have a geodesic with such a dependence, we have to set the value of $\eta$ different from $0$. If we set $\eta = \pi/2$, which is the case we analyzed in section \ref{Point-like-string}, then we have a degenerate solution since the determinant of the metric of the pp-wave background is zero. This fact implies that we lose one of the coordinates and the pp-wave backround becomes 9-dimensional, a feature that is undesirable. On the other hand, if  we keep the value of $\eta$ fixed but arbitrary, then in order for the geodesic ansatz to be consistent we have to fix the value of 
$\omega$ in terms of the deformation parameters. In this way we obtain a pp-wave solution, but with the cost that it is extremely complicated, e.g. it is impossible to bring it in the Brinkmann form and study the spectrum of bosonic excitations. For all these reasons we choose to expand around $\eta = 0$. 

In particular, we expand the coordinates in the following way
\begin{equation}
 \begin{aligned}
  & t = \kappa \, U \, , \quad  v = \m^2\, m \, U -  \frac{\om}{m} \, \frac{y_2}{R} \, , \quad 
  \chi = \om \, U + \frac{y_2}{R} + \frac{1}{\om} \, \frac{V}{R^2} \, ,
  \\[5pt]
  & z = \sqrt{\frac{\kappa}{m}} \, \Big( 1 + \frac{y_1}{R}  \Big) \, , \quad 
  x_1 = \sqrt{\frac{\kappa}{m}} \, \frac{y_3}{R} \, , \quad 
  x_2 = \sqrt{\frac{\kappa}{m}}  \, \frac{y_4}{R}  \quad  \& \quad \eta = \frac{r}{R} \, .
 \end{aligned}
\end{equation}
One can check that the dominant terms of the previous expansion in the $R \rightarrow \infty$ limit guarantee 
that the null condition for the geodesic is satisfied, provided that the parameters $\kappa$, $\om$ and $m$ are related through the following dispersion relation
\begin{equation}
\kappa^2 = \om^2 + m^2 \m^2 \, .
\end{equation}
Notice that although this dispersion relation is $\gamma$ independent the spectrum of the pp-wave background will depend on $\gamma$ in a non-trivial way. 

Using the aforementioned expansion for the coordinates and taking the limit $R \rightarrow \infty$ in the line element,
we end up with the following expression
\begin{equation}
 \label{ppwaveMetric}
 \begin{aligned}
  & ds^2_{pp} = 2 \, dU \, dV + \Bigg[ 2 \, \om \Big( y_1 \, dy_2 - y_2 \, dy_1 + 
  y_5 \, dy_6 - y_6 \, dy_5 + y_7 \, dy_8 - y_8 \, dy_7 \Big)
 \\[5pt]
  & \qquad + \m \, m \, \gamma \, \Big( y_6 \, dy_5 - y_5 \, dy_6 + y_7 \, dy_8 - y_8 \, dy_7 \Big)\Bigg] dU 
  + \sum\limits_{i = 1}^8 dy^2_i  - F \, dU^2 \, ,
  \\[5pt]
  & {\rm where} \quad F = 4 \, \mu^2 \, m^2 \,  y^2_1 + \big( \mu^2  \, m^2 + \om^2 \big) \big( y^2_3 + y^3_4 \big) - \frac{\g^2}{4} \,  \big( \m^2 \, m^2 - \om^2 \big)  \sum\limits_{i = 5}^8 y^2_i \, .
 \end{aligned}
\end{equation}
Moreover, in order to arrive to the previous expression for the metric of the pp-wave we considered the following change of coordinates
\begin{equation}
 y_5 + i \, y_6 \, = \, r \, \sin\frac{\th}{2} \, e^{i \frac{\phi + \psi}{2}} 
 \quad \& \quad  
 y_8 + i \, y_7 \, = \, r \, \cos\frac{\th}{2} \, e^{i \frac{\phi - \psi}{2}}
\end{equation}
and shifted $V$ as follows: $V \rightarrow V - \om \, y_1 \, y_2$.
The geometry \eqref{ppwaveMetric} has a vanishing Ricci scalar (i.e.  $R_\mu^{\,\,\,\mu} = 0)$, as it should be for a pp-wave geometry.

We would like to further simplify the expression for the line element of our pp-wave background and bring it into the Brinkmann form. 
In order to do this we perform three rotations in the three complex planes defined by the coordinate pairs $(y_1, y_2)$, 
$(y_5, y_6)$ and $(y_7, y_8)$ as it is described in the following relations
\begin{equation}
\begin{aligned}
 & y_1 + i \, y_2 \, = \, \left(\tilde{y}_1 + i \, \tilde{y}_2 \right) \, e^{- i \, \om \, U} \, , \quad 
  y_5 + i \, y_6 \, = \, \left(\tilde{y}_5 + i \, \tilde{y}_6\right) \, e^{i \, \frac{\mu \, m \,  \g - 2 \, \om}{2} \, U} \, ,
  \\[5pt]
  & \qquad \qquad \qquad y_8 + i \, y_7 \, = \,  \left(\tilde{y}_8 + i \, \tilde{y}_7 \right) \, e^{i \, \frac{\mu \, m \, \g + 2 \, 
  \om}{2} \, U} \, .
\end{aligned}
\end{equation}
After dropping the tildes, the geometry has the standard line element
\begin{equation}  \label{ds2ppBrinkmann}
ds^2_{pp} \, = \, 2 \, dU \, dV + \sum\limits_{i = 1}^8 dy^2_i - H \, dU^2 
\end{equation}
and the function $H$ is defined as follows
\begin{eqnarray}  \label{Hdef-Brinkmann}
H &= &\om^2 \big( y^2_1 + y^2_2 \big) + \big( \m^2 \, m^2 + \om^2 \big) \big( y^2_3 + y^2_4 \big) + 
\frac{\om}{4} \Big[ \big( 4 + \g^2 \big) \, \om - 4 \,\mu \, m  \, \g \Big] \big( y^2_5 + y^2_6 \big) 
\nonumber \\[5pt]
&+& \frac{\om}{4} \Big[ \big( 4 + \g^2 \big) \, \om + 4 \, \mu \,  m \, \g \Big) \big( y^2_7 + y^2_8 \big) + 
4 \, \m^2 \, m^2 \Big[ y_1 \,\cos(\om \, U) + y_2 \, \sin(\om \, U) \Big]^2 \, .
\end{eqnarray}
The geometry is supported by a NS $2-$form and a RR $5-$form 
\begin{equation}
 \label{B2ppBrinkmann}
    \begin{aligned}
     B_2 & =\m \, m \, dU \wedge \Big(y_2 \, dy_1 - y_1 \, dy_2 + y_5 \, dy_6 - y_6 \, dy_5 + y_7 \, dy_8 - y_8 \, dy_7\Big)
     \\[5pt]
     & + \frac{\g \, \om}{2} dU \wedge \Big( y_6 \, dy_5 - y_5 \, dy_6 + y_7 \, dy_8 - y_8 \, dy_7\Big) \, ,
     \\[5pt]
     F_5 & = 4 \, \om \, dU \wedge \Big(dy_1 \wedge dy_2 \wedge dy_3 \wedge dy_4 - dy_5 \wedge dy_6 \wedge dy_7 \wedge dy_8\Big) \, ,
    \end{aligned}
\end{equation}
where in the expression for the $2-$form we have dropped terms that are total derivatives. Notice that if we set $\gamma=0$ the expressions  \eqref{ds2ppBrinkmann}, \eqref{Hdef-Brinkmann} and \eqref{B2ppBrinkmann} reduce to the 
expressions of \cite{Georgiou:2019lqh}. Moreover if we set  $\mu=0$, we obtain the pp-wave solution that was 
analyzed in \cite{Niarchos:2002fc, Lunin:2005jy} (see also \cite{Avramis:2007wb}).  

To calculate the spectrum of closed strings propagating on the pp-wave background that is given 
\eqref{ds2ppBrinkmann} and \eqref{B2ppBrinkmann}, we have to write down the bosonic string action and derive the equations of motion. To set-up the conventions, in appendix \ref{String_EOM} we have collected, the Polyakov action %, the 
%corresponding equations of motion for the string 
 and the Virasoro constrains. %Below we will apply this machinery on the pp-background that we have described earlier in the current section.   

Choosing the light-cone gauge (i.e. $U = \a' p_+ \tau $), the Virasoro constraints will determine $V$ in terms of the remaining eight degrees of freedom.  
It turns out that the equations of motion for the scalars $y_3$ and $y_4$ are completely decoupled, while the rest couple in pairs. Below we analyze them separately.

%%%%%%%%%%%%%%%%%%%%%%%%%%%%%%%%%%%%%%%%%%%%%%%%%%%%%%%%%%%%

\subsubsection*{Equations of motion for $y_1 \, \& \, y_2$}

The equations of motion for the scalars $y_1$ and $y_2$ have a more complicated form due to the explicit dependence on $\tau$
\begin{equation} \label{EOM_y1y2}
 \begin{aligned}
  & \Box \, y_1 + 2 \, \a' \, p_+ \, \mu \, m \, \partial_\s y_2 
  \\[5pt]
  & - \a'^2 \, p_+^2 \Bigg[ \om^2 y_1 - 4 \, \m^2 m^2 \cos\big( \a' p_+ \om \tau \big) \, 
  \Big[  y_1 \cos\big( \a' p_+ \om \tau \big) + y_2 \sin\big( \a' p_+ \om \tau \big) \Big]  \Bigg] = 0 \, ,
  \\[5pt]
  & \Box \, y_2 - 2 \, \a' \, p_+ \, \mu \,  m \, \partial_\s y_1 
  \\[5pt]
  & - \a'^2 \, p_+^2 \Bigg[ \om^2 y_2 - 4 \, \m^2 m^2 \sin\big( \a' p_+ \om \tau \big) \, 
  \Big[  y_1 \cos\big( \a' p_+ \om \tau \big) + y_2 \sin\big( \a' p_+ \om \tau \big) \Big]  \Big] = 0 \, .
 \end{aligned}
\end{equation}
Notice here that this set of equations does not depend on the $\gamma$ deformation parameter and the analysis has been 
presented in \cite{Georgiou:2019lqh}. However, in order to be complete we briefly repeat it here. 
To eliminate the dependence in $\tau$, we perform the following rotation
\begin{equation}
 y_1 = \tilde{y}_1 \cos\big( \a' p_+ \om \tau \big) - \tilde{y}_2 \sin\big( \a' p_+ \om \tau \big) \, , \quad y_2 = \tilde{y}_1 \sin\big( \a' p_+ \om \tau \big) + \tilde{y}_2 \cos\big( \a' p_+ \om \tau \big) \, .
\end{equation}
an the system of equations in \eqref{EOM_y1y2} becomes
\begin{equation}
 \begin{aligned}
  & \Box \, \tilde{y}_1 - 4 \, \a'^2 \, p_+^2 \, \m^2 \, m^2 \, \tilde{y}_1 + 2 \, \a' \, p_+ 
  \big(  \om \, \partial_\tau \tilde{y}_2 + \m \, m \, \partial_\s \tilde{y}_2 \big) \, = \, 0 \, ,
  \\[5pt]
  & \Box \, \tilde{y}_2 - 2 \, \a'  \, p_+ \big(  \om \, \partial_\tau \tilde{y}_1 + \m \, m \, \partial_\s \tilde{y}_1 \big) \, = \,  0 \, .
 \end{aligned}
\end{equation}
Imposing the ansatz
\begin{equation}
 \tilde{y}_1 = \sum\limits_{n = - \infty}^{\infty} \a_n e^{- i \tilde{\om}_n \tau + n \s} \, , \qquad \tilde{y}_2 = \sum\limits_{n = - \infty}^{\infty} \b_n e^{- i \tilde{\om}_n \tau + n \s} \, ,
\end{equation}
with $\a_n$ and $\b_n$ being constants, we end up with a homogeneous algebraic system for $\a_n$ and $\b_n$. Requiring that the determinant of the matrix that defines the aforementioned algebraic system with respect to $\a_n$ and $\b_n$ vanishes, we obtain a quartic equation for the eigenfrequencies $\tilde{\om}_n$
\begin{equation}
 \tilde{\om}^4_n + \g_2 \, \tilde{\om}^2_n + \g_1 \, \tilde{\om}_n + \g_0 = 0 \, ,
\end{equation}
with
\begin{equation}
 \g_0 = n^4 \, , \quad \g_1 = 8 \,  \a'^2 \, p_+^2 \, \m \, m \, \om \, n  \quad  \& \quad 
 \g_2 = - 2 \, n^2 - 4 \, \a'^2 \, p_+^2 \big(  \om^2 + \m^2 \, m^2 \big) \, .
\end{equation}
The solution of the quartic equation above is
\begin{equation}  \label{om12}
 \tilde{\om}_n^{(1)} = - S \pm \frac{1}{2} \sqrt{- 4 S^2 - 2 \g_2 + \frac{\g_1}{S}} \, , \qquad \tilde{\om}_n^{(2)} = S \pm \frac{1}{2} \sqrt{- 4 S^2 - 2 \g_2 - \frac{\g_1}{S}} \, ,
\end{equation}
where
\begin{equation}
 \begin{aligned}
  & S = \frac{1}{2} \sqrt{- \frac{2}{3} \g_2 + \frac{1}{3} \Big(  Q + \frac{\D_1}{Q} \Big)} \, , \qquad Q^3 = \frac{1}{2} \Big(  \D_1 + \sqrt{\D_1^2 - 4 \D_0^3} \Big) \, ,
  \\[5pt]
  & {\rm with } \quad \D_1 = 2 \, \g_2^3 + 27 \, \g_1^2 - 72 \, \g_2 \, \g_0 \quad \& \quad  \D_0 = \g_2^2 + 12 \, \g_0 \, .
 \end{aligned}
\end{equation}

%%%%%%%%%%%%%%%%%%%%%%%%%%%%%%%%%%%%%%%%%%%%%%%%%%%%%%%

\subsubsection*{Equations of motion for $y_3 \, \& \, y_4$}

The decoupled equations for the scalars $y_3$ and $y_4$ do not depend on the $\gamma$ deformation parameter and read
\begin{equation}
 \Box y_i - \a'^2 p_+^2 \, \k^2 y_i = 0 \, , \qquad i = 3, 4 \, .
\end{equation}
These can be easily solved using the plane wave ansatz $y_i \sim e^{- i \,  \om_{i;n} \, \tau + i \, n \, \s}$ which implies that the eigenfrequencies for $y_i$ are
\begin{equation}  \label{om34}
 \om^2_{i;n} = \a'^2 \, p_+^2 \, \k^2 + n^2 = \a'^2 \, p_+^2 \, \big(  \om^2 +\m^2 \, m^2  \big) + n^2 \, , \qquad i = 3, 4 \, .
\end{equation}

%%%%%%%%%%%%%%%%%%%%%%%%%%%%%%%%%%%%%%%%%%%%%%%%%%%%%%

\subsubsection*{Equations of motion for $y_5 \, \& \, y_6$}

We now move to the equations for the scalars $y_5$ and $y_6$, which are coupled due to the derivative terms
\begin{equation}
 \label{eomsy5y6}
 \begin{aligned}
  & \Box \, y_5 - \a'^2 \, p_+^2 \, \om \left[ \left( 1 + \frac{\g^2}{4} \right) \om - \m \, m \, \g  \right] y_5 + 
  \a' \, p_+ \big(  \g \, \om - 2 \, \m \, m \big) \partial_\s y_6 = 0 \, ,
  \\[5pt]
  & \Box \, y_6 - \a'^2 \, p_+^2 \, \om \left[ \left( 1 + \frac{\g^2}{4} \right) \om - \m \, m \, \g  \right] y_6 - 
  \a'  \, p_+ \big(  \g \, \om - 2 \, \m \, m \big) \partial_\s y_5 = 0 \, .
 \end{aligned}
\end{equation}
The above set of equations can be decoupled if we consider the linear combinations $y^\pm_{56} = y_5 \pm i y_6$. Then for the $y^\pm_{56}$ we get
\begin{equation}
 \begin{aligned}
  & \Box y^+_{56} - \a'^2 p_+^2 \om \left[ \left( 1 + \frac{\g^2}{4} \right) \om - \m \, m \, \g  \right]  y^+_{56} - 
  i \, \a' \, p_+ \big(  \g \, \om - 2 \, \m \, m \big) \partial_\s y^+_{56} = 0 \, ,
  \\[5pt]
  & \Box y^-_{56} - \a'^2 p_+^2 \om \left[ \left( 1 + \frac{\g^2}{4} \right) \om - \m \, m \, \g  \right] y^-_{56} + 
  i \, \a' \, p_+ \big(  \g \, \om - 2\,  \m \, m \big) \partial_\s y^-_{56}= 0 \, .
 \end{aligned} 
\end{equation}
Choosing again a plane-wave ansatz $y^\pm_{56} \sim e^{- i \om^\pm_{56;n} \tau + i n \s}$ we get
\begin{equation}
 \label{om56pm}
 \big(  \om^\pm_{56;n} \big)^2 = \a'^2 \, p_+^2 \, \om \left[ \left( 1 + \frac{\g^2}{4} \right) \om - \m \, m \, \g  \right] \pm 
 n \, \a' \, p_+ \big(  2 \, \m \, m - \g \, \om \big) + n^2 \, .
\end{equation}
It is important to notice that the frequencies depend on both deformation parameters $\mu$ and $\gamma$ in a non-trivial way.
%%%%%%%%%%%%%%%%%%%%%%%%%%%%%%%%%%%%%%%%%%%%%%%%%%%%%%%%

\subsubsection*{Equations of motion for $y_7 \, \& \, y_8$}

The equations for the scalars $y_7$ and $y_8$ can be treated similarly to those for $y_5$ and $y_6$. In particular we have that
\begin{equation}
 \label{eomsy7y8}
 \begin{aligned}
  & \Box \, y_7 - \a'^2 \, p_+^2 \, \om \left[ \left( 1 + \frac{\g^2}{4} \right) \om + \m \, m \, \g  \right]  y_7 - 
  \a' \, p_+ \big(  \g \, \om + 2 \, \m \, m \big) \partial_\s y_8 = 0 \, ,
  \\[5pt]
  & \Box \, y_8 - \a'^2 \, p_+^2 \, \om \left[ \left( 1 + \frac{\g^2}{4} \right) \om + \m \, m \, \g  \right] y_8 + 
  \a' \, p_+ \big(  \g \, \om + 2\,  \m \, m \big) \partial_\s y_7 = 0 \, .
 \end{aligned}
\end{equation}
Notice that these can be obtained from the equations of motion for $y_5$ and $y_6$ in \eqref{eomsy5y6} 
by sending $y_5 \mapsto y_8$, $y_6 \mapsto y_7$ and $\m \mapsto - \m$. Thus the eigenfrequencies for the decoupled scalars $y^\pm_{78} = y_8 \pm i y_7$ can be obtained from those for $y^\pm_{56}$ in eq. \eqref{om56pm} simply by sending $\m \mapsto - \m$, i.e.
\begin{equation}
 \label{om78pm}
 \big(  \om^\pm_{78;n} \big)^2 = \a'^2 \, p_+^2 \, \om \left[ \left( 1 + \frac{\g^2}{4} \right) \om + \m \, m \, \g  \right] \mp 
 n \, \a'  \, p_+ \big(  2 \, \m \, m + \g \, \om \big) + n^2 \, .
\end{equation}
Notice that in the case of $\gamma=0$ we obtain the frequencies of the Schr\"{o}dinger background that were computed in \cite{Georgiou:2019lqh}, while for $\mu=0$ the frequencies in \eqref{om12}, \eqref{om34}, \eqref{om56pm} and
 \eqref{om78pm} reduce to those computed in \cite{Niarchos:2002fc, Lunin:2005jy} (see also \cite{Avramis:2007wb}). 

Moreover, in the pp-wave background, one can try to construct a Giant Graviton solution in the same way that we did in section \ref{Giant-Graviton}. The bottomline of the construction is that this pp-wave graviton, in the same way as the graviton of section \ref{Giant-Graviton}, will be independent of the $\gamma$ deformation parameter. As a result such a computation reduces to the computation of  \cite{Georgiou:2020qnh}. In order to obtain a Giant Graviton that depends on the $\gamma$ deformation parameter, one should try for a different geodesic. 

\subsection{PP-wave spectrum and dispersion of the giant magnon}
\label{spectrum-up}

In this section, and based on the bosonic spectrum on the pp-wave background derived in the
previous section, we make an educated guess for the {\it exact} in $\lambda$ anomalous dimensions of the operators that are dual to the string excitations \eqref{om34}, \eqref{om56pm} and \eqref{om78pm}. To this end we closely follow section 4 of \cite{Georgiou:2019lqh}.

One starts from the fact that the light-cone string Hamiltonian 
is linear in the generators $E$, $M$ and $J$,
namely,
\begin{eqnarray} \label{Ham0}
H_{\rm l.c.}  \equiv  p^-  =  p_+  =  i  \frac{\partial}{\partial x^+}  =
i  \Bigg[\k  {\partial \ov \partial T}   +  \om   {\partial \ov \partial \chi} +  \mu^2  m  {\partial \ov \partial V} \Bigg] \, .
\end{eqnarray}
Furthermore, since the conserved charges in the original Schr\"{o}dinger background are given by
\begin{eqnarray} \label{charges}
E  =  i \, {\partial \ov \partial T}\, , \quad
J  =  - \, i \, {\partial \ov \partial \chi} \ , \quad
M \, = \, - \, i  \, {\partial \ov \partial V}\, ,
\end{eqnarray}
one gets the following relation for the light-cone Hamiltonian in terms of the conserved charges of the original doubly deformed spacetime
\begin{eqnarray} \label{Ham1}
H_{\rm l.c.} =  \k  E  - \om  J  -  \mu^2  m  M\, .
\end{eqnarray}
Additionally, the light-cone momentum is given by
\begin{eqnarray} \label{lc-moment}
 \a'  p^+ =  \a'  p_-  = -  i  \a' \frac{\partial}{\partial x^-}  = 
 -  i {\a' \ov \om R^2} {\partial \ov \partial \chi} =  {J\ov \sqrt{\l}} \, .
\end{eqnarray}
In \eqref{lc-moment}, in order to simplify the notation, we have set $\om=1$ and we have also used the relation between
the radius $R$ and the 't Hooft coupling $\l$, that is $R^2/\a'=\sqrt{\l}$.

By identifying the light-cone Hamiltonian \eqref{Ham1} with the energy of a single string
excitation given by \eqref{om78pm}, that is $H_{\rm l.c.}= \om^\pm_{78;n}$,
one  gets that
\begin{eqnarray} \label{disp0}
{\k \, \sqrt{\l}  E -  \mu^2  m  \sqrt{\l} M \ov J}  - \sqrt{\l}  = 
\sqrt{1+{\gamma^2\ov 4}+\mu m \gamma+{n^2 \l \ov J^2}  \mp  {  n  (2\mu  m+\gamma)  \sqrt{\l} \ov J}} \ .
\end{eqnarray}
 The next step is to write the
angular momentum, energy and particle number of the string moving on the pp-wave geometry as the corresponding
quantities of the point-like BMN particle plus corrections of order $\mathcal O(\l^0)$, that is
\begin{eqnarray} \label{expansion}
E  = \sqrt{\l}  \k  +  \epsilon_1 \ , \quad
J  =  \sqrt{\l}  +  j_1 \ ,\quad
M  =  \sqrt{\l}  m +  m_1 \ .
\end{eqnarray}
One then solves \eqref{expansion}  for the quantities $\sqrt{\l} \k$, $\sqrt{\l} $ and $\sqrt{\l} m$ in terms of $E$, $J$
and $M$ and substitute the resulting expressions in \eqref{disp0} to obtain
\be
\label{disp1}
{E^2  -  \mu^2  M^2 \ov J}  -  J -Y_1 = 
\sqrt{1+{\gamma^2\ov 4}+\mu m \gamma+{n^2 \l \ov J^2}  \mp  {  n  (2\mu  m+\gamma)  \sqrt{\l} \ov J}}  \ ,
\ee
where $Y_1$ collects the terms involving $\e_1,j_1$ and $m_1$ and is given by
\be
Y_1 =  {\epsilon_1  E  -  \mu^2 M  m_1 \ov J}  -  j_1 \, .
\ee
The last step is to  express $J$ in terms of $E$, $\epsilon_1$, $M$, $m_1$ and $j_1$.
This  can be done by plugging \eqref{expansion} in the holding relation $\kappa^2=\mu^2 m^2+\omega^2$ to get
\begin{eqnarray} \label{inter}
%\begin{split}
&& E^2  -  \mu^2  M^2 -  J^2  =  Z \quad {\rm with}
\nonumber \\[5pt]
&& Z = 2 \left(\epsilon_1  \k \sqrt{\l}  - \sqrt{\l} j_1  - \mu^2  m \sqrt{\l} m_1\right)  + \epsilon_1^2  -  j_1^2  -  \mu^2  m_1^2 \ .
%\end{split}
\end{eqnarray}
The first equation in \eqref{inter} can now be solved for $J$ to give
\begin{equation} \label{Jay}
J = \sqrt{E^2  - \mu^2 M^2} \Bigg[1  -  {1\ov 2}  {Z\ov E^2 - \mu^2 M^2 }\Bigg] \, ,
\end{equation}
where in order to derive this last equation we have approximated $\sqrt{1-{Z\ov E^2 -\mu^2 M^2 }}$ by
$1-{1\ov 2}{Z\ov E^2-\mu^2 M^2 }$, since $Z$ is of ${\cal O}(\sqrt{\l})$ while
$ E^2-\mu^2 M^2$ is of ${\cal O}(\l)$.

The result for $J$ should be now substituted in \eqref{disp1} to give
\begin{equation} \label{disp-fin}
\sqrt{ E^2  -  \mu^2  M^2 }  -  J  = 
\sqrt{1+{\gamma^2\ov 4}+\mu m \gamma+{n^2 \l \ov J^2}  \mp  {  n  (2\mu  m+\gamma)  \sqrt{\l} \ov J}} +  {\cal W}\ ,
\end{equation}
with
\begin{equation} \label{def-W}
{\cal W} = {1 \ov 2 \, J} \, \left(\epsilon_1^2 \, - \, j_1^2 \, - \, \mu^2 \, m_1^2\right) \, + \,
\frac{Z^2}{4 \, J \, \left(Z + J^2 \right)} \, .
\end{equation}
Let us stress that  ${\cal W} $ in \eqref{disp-fin} should be ignored since it scales as ${1/J}$ and it becomes zero
in the strict $J$ infinity limit in which the giant magnon is defined.
%In addition, if ones wishes to compare \eqref{disp-fin} with the dispersion relation of the giant magnon \eqref{dispersion}
%in the original background one should also ignore the third term in the square root of the right hand side of \eqref{disp-fin}.
%The reason is that in the large $\l$ limit the aforementioned term scales as $\sqrt{\l}$ and is thus suppressed
%with respect to the previous term under the square root. In order to find this contribution in the original Schr\"{o}dinger background one should calculate the $\a'$ corrections to the dispersion relation \eqref{dispersion}

%Finally, one may perform the analysis of this section for the eigenfrequency \eqn{omega3}. The result is again in agreement with the dispersion relation of the giant magnon in the original Schr\"{o}dinger background \eqref{dispersion}. We, thus, see that the BMN spectrum carries more refined information compared to the  strong coupling result of \eqref{dispersion}.
In precisely the same way one obtains the following exact in $\l'={\l \ov J^2}$ dispersion relation for the excitations of  \eqref{om56pm}.
\begin{equation} \label{disp-fin-2}
\sqrt{ E^2  -  \mu^2  M^2 }  -  J  = 
\sqrt{1+{\gamma^2\ov 4}-\mu m \gamma+{n^2 \l \ov J^2}  \pm  {  n  (2\mu  m-\gamma)  \sqrt{\l} \ov J}} \ .
\end{equation}
Finally, for the excitations of \eqref{om34} which do not depend at all on $\gamma$ one straightforwardly obtains
\begin{equation} \label{disp-fin-3}
\sqrt{ E^2  -  \mu^2  M^2 }  -  J  = 
\sqrt{1+\m^2 m^2+{n^2 \l \ov J^2} } \ .
\end{equation}

We close this section by making an educated guess for the {\it exact} in $\lambda$ dispersion relation of the magnon excitations in the original doubly deformed 
background. Indeed, inspired by the form of the pp-wave spectrum it is plausible to conjecture that the exact dispersion relation corresponding to \eqref{disp-fin-3} is given by
%\footnote{It may of course be the case in which non-trivial interpolating functions are needed between the weak and string coupling results.}
%
\begin{eqnarray} \label{dispersion-1}
\sqrt{E^2-\mu^2 M^2}-J  =  \sqrt{1+\mu^2 m^2+\frac{\lambda}{\pi^2} \sin^2\frac{p}{2}} \, ,
\quad J \rightarrow \infty\ ,
\end{eqnarray}
while the ones corresponding to \eqref{disp-fin-2} and \eqref{disp-fin} are given by (for $J \rightarrow \infty$)
\begin{eqnarray} \label{dispersion-3}
\sqrt{E^2-\mu^2 M^2}-J =  \sqrt{1+{\gamma^2\ov 4}-\mu m \gamma+\frac{\lambda}{\pi^2} \sin^2\frac{p}{2}\pm  \left(\mu m-{\gamma\ov 2}\right) \frac{ \sqrt{\l}}{\pi} \, \sin{p}}
\end{eqnarray}
and
\begin{eqnarray} \label{dispersion-2}
\sqrt{E^2-\mu^2 M^2}-J =  \sqrt{1+{\gamma^2\ov 4}+\mu m \gamma+\frac{\lambda}{\pi^2} \sin^2\frac{p}{2}\mp  \left(\mu m+{\gamma\ov 2}\right) \frac{ \sqrt{\l}}{\pi} \, \sin{p}}
\end{eqnarray}
respectively.

Indeed, for small values of the momentum $p=\frac{2\pi  n}{J}, \,\,\,n\in Z$ one gets  \eqref{disp-fin-3}, \eqref{disp-fin-2} and \eqref{disp-fin}  respectively.
It would be interesting to identify the corresponding field theory operators and check whether their exact in $\lambda$ conformal dimensions
are indeed given by \eqref{dispersion-1}, \eqref{dispersion-3} and \eqref{dispersion-2}.
It is important to note that these dispersion relations can be rewritten in a form relevant for the dual null dipole doubly deformed CFT of section \ref{FT} by using the relations $\mu = \frac{\sqrt{\lambda}}{2 \pi}  L$ and $\gamma=\frac{\sqrt{\lambda}}{4 \pi}  \tilde \gamma$ (see \cite{Guica:2017mtd}), where $\tilde L$ and $\tilde \g$ are the parameters entering the star product that deforms the parent theory, ${\cal N}=4$ SYM. Having done this, equations \eqref{dispersion-1}, \eqref{dispersion-3} and \eqref{dispersion-2} have the correct weak coupling expansion in integer powers of $\lambda$. A final comment is in order. Note that for $\gamma=0$ and $\l\gg 1$ the dispersion relations \eqref{dispersion-1}, \eqref{dispersion-3} and \eqref{dispersion-2} reduce to the dispersion relation of the giant magnon solution of \cite{Georgiou:2017pvi}, while for $\mu=0$ and $\l\gg 1$ to the dispersion relation of the giant magnon solution of \cite{Chu:2006ae}.

%%%%%%%%%%%%%%%%%%%%%%%%%%%%%%%%%%%%%%%%%%%%%%%%%%%%%%%%%%%%%%%%%%%%%%

\section{Conclusions}
\label{concl}
In this work we performed a thorough and detailed study of a new example of the AdS/CFT correspondence, namely of the marginally deformed Schr\"{o}dinger background. From the gravity side this solution is constructed by performing  a series of TsT transformations along two of the five-sphere isometries of the Schr\"{o}dinger solution. As a result the background depends on two deformation parameters and is integrable for large number of colours $N$. From the field theory side, the Lagrangian of the theory is obtained from that of $\mathcal {N}=4$ SYM by introducing the appropriate star product among the fields. The resulting gauge theory is integrable at the planar level.

Putting an emphasis on the study of the correspondence for this doubly-deformed background, we identified a point-like string solution that depends on both deformation parameters and we derived its dispersion relation. 
Exploiting the power of the correspondence, we identified the dual field theory operators and using the Landau-Lifshitz coherent state approach, we managed to reproduce the leading terms of the aforementioned dispersion relation. This is a highly non-trivial test of the correspondence. We performed the computation at leading order, in which case the two deformations are disentangled. It would be very interesting to extend this calculation beyond the leading order, that would include terms depending on products of the two deformations simultaneously. In this case the corresponding Hamiltonian wouldn't be the sum of two terms, one related to $AdS_3$ and the other to $S^3$, as it happens with the one loop Hamiltonian \eqref{Ham}.

Moving forward to the study of field theory observables through their gravity realisation, we focused our attention on the evaluation of the Wilson loop expectation value. The Wilson loop consists of two straight lines extending along the time direction and sitting at two points in space separated by a distance $\mathbb{L}$. We calculate the energy of this configuration as a function of the separation length.
This corresponds, as usual, to the potential between two quarks sitting at a distance  $\mathbb{L}$ on the boundary \cite{Maldacena:1998im}. We identified an embedding ansatz for the string that depends on both deformation parameters and calculated the energy of the aforementioned configuration as a function of the separation boundary length. The behaviour of the energy exhibits interesting properties: 
For values of the length much smaller than the Schr\"{o}dinger parameter, the energy is inversely proportional to the $\mathbb{L}$, a behaviour that resembles that of a conformal theory. 
When the separation length is much less than the Schr\"{o}dinger parameter, an intriguing confining behaviour arises. The energy is linearly proportional to the length with the constant of proportionality being inversely proportional to $\mu^2$. As $\mu \to 0$, we obtain the expected conformal behaviour for all the values of the length. Notice here that the observations we detailed above hold for every value of the deformation parameter $\gamma$. Essentially the presence of $\gamma$ is only ``dressing" the results, but does not alter their qualitative characteristics or the physical interpretation.

The next observable we shed the light on was the Giant Graviton solution. Even if both deformation parameters appear in the induced metric of the D3-brane action, a cancellation takes place and the marginal deformation $\gamma$ disappears from the energy of the Giant Graviton. It would be very interesting to perform a stability analysis around the aforementioned solution. Based on similar cases in the literature, the vibration modes will depend on both deformation parameters $\mu$ and $\gamma$. Examining those modes a relation between the parameters might arise, in order for the solution to be stable.  

It is well-known that string theory is solvable in the pp-wave limit. Furthermore, the spectrum of the strings on the pp-wave background provides information about the dimensions of a particular subsector of gauge theory operators having large charges. Having these in mind we took the Penrose limit of the marginally deformed  Schr\"{o}dinger background along a certain null geodesic. This geodesic was special, in the sense that the final background has a dependence on both deformation parameters. 
To elaborate on the pp-wave solution we have calculated the spectrum of closed bosonic strings in the deformed background. The eigenfrequencies of the eight transverse physical degrees of freedom give a prediction for the anomalous dimensions of the dual field theory operators as an exact function of the effective coupling $\lambda' ={\lambda \over J^2}$. Four of the eigenfrequencies depend only on the  Schr\"{o}dinger parameter while the remaining four depend on both the marginal and the  Schr\"{o}dinger parameters. Finally, based on the string spectrum on the pp-wave geometry, we made an educated guess for the {\it exact} in $\lambda$ dispersion relation of the magnon excitations in the original doubly deformed background which provides us with an exact prediction for the dimensions of the dual field theory operators.

The analysis we performed so far leaves a number of interesting open questions, besides the ones we have already discussed in the process of summarizing our findings. An important calculation that can be done is the construction of the Giant Magnon solution and the corresponding computation of its dispersion relation. With the proper ansatz this relation would depend on both deformation parameters. Notice that while in the present work the dispersion relation on the pp-wave background is exact in $\lambda'$ with the momentum of the magnon being $p\ll1$, the dispersion relation of the Giant Magnon will be in the strong coupling regime $\lambda \gg 1$ with the momentum excitation being finite. While the two computations are independent and different, there is a region of the parametric space for which there is an overlap. In section \ref{spectrum-up}, we made an informed guess for the exact in $\lambda$ dispersion relation of the giant magnon in the background \eqref{metric}, \eqref{dilaton} and \eqref{def-B-field}. It would be interesting to identify the exact form of the field theory operators and calculate their anomalous dimension in order to test these dispersion relations at the weak coupling regime. 
Along the same lines, one may try to employ, after generalising, the method of \cite{Guica:2017mtd} that is based on the Q-Baxter equation in order to calculate the spectrum of our doubly deformed background. This is necessary since, like in \cite{Guica:2017mtd}, the corresponding to \eqref{Lagrangian-FT}  spin chain does not have a reference state, rendering thus the usual Bethe ansatz approach inapplicable.  

Another interesting direction is the calculation of the spectra of massless excitations in the marginally deformed  Schr\"{o}dinger background. One needs to solve the Laplace equation in the deformed background and the computation is performed by transforming the associated differential equations into Schr\"{o}dinger problems. 
Within the correspondence, those fluctuations correspond to certain operators in the gauge theory.

%%%%%%%%%%%%%%%%%%%%%%%%%%%%%%%%%%%%%%%%%%%%%%%%%%%%%%%%%%%

\section*{Acknowledgments}
The work of G.G. and D.Z. has received funding from the Hellenic Foundation for Research and Innovation (HFRI) and the General Secretariat for Research and Technology (GSRT), under grant agreement No 15425. The research work of G.I. is supported by the Einstein Stiftung Berlin via the Einstein International Postdoctoral Fellowship program ``Generalised dualities and their holographic applications to condensed matter physics'' (project number IPF-2020-604). G.I. is also supported by the Deutsche Forschungsgemeinschaft (DFG, German Research Foundation) via the Emmy Noether program ``Exploring the landscape of string theory flux vacua using exceptional field theory'' (project number 426510644).

%%%%%%%%%%%%%%%%%%%%%%%%%%%%%%%%%%%%%%%%%%%%%%%%%%%%%%%%

\appendix

\section{The non-supersymmetric Schr\"{o}dinger solution}
\label{Schrodinger}

We present the type-IIB solution \cite{Guica:2017mtd} which is obtained after applying the appropriate TsT transformation to the $AdS_5 \times S^5$ solution. The background is non-supersymmetric and has the Schr\"{o}dinger group as its symmetry group. 

The background consists of a metric with line element:
\begin{equation}
 \begin{aligned}
  R^{-2} \, ds^2 & = - \left[ 1 + \frac{\m^2}{z^4} + \frac{x^2_1 + x^2_2}{z^2} \right] dt^2 + \frac{1}{z^2} 
  \left(  2 \, dt dv + dx^2_1 + dx^2_2 + dz^2 \right)
  + d\chi^2 + d\eta^2 
  \\[5pt]
  & + \sin^2\eta \, d\chi \big(  d\psi - \cos\th \, d\phi \big) + \frac{1}{4} \, \sin^2 \eta \left(  d\th^2 + d\phi^2 + d\psi^2 -2 \, \cos\th \, d\phi \, d\psi \right) \, ,
 \end{aligned}
\end{equation}
where $\m$ is a deformation parameter. There is also a non-trivial NS two-form given below:
\begin{equation}
 B_2 = \frac{R^2 \, \m}{z^2} \, dt \wedge \big( d\chi + \om \big) 
 \quad {\rm with} \quad 
 \om = \frac{1}{2} \sin^2\eta \big(  d\psi - \cos\th \, d\phi \big) \, .
\end{equation}
Finally, the RR sector contains only the self-dual five-form which is:
\begin{equation}
 F_5 = \frac{R^4}{2} \big(1 + \star \big) \cos\eta \, \sin^3\eta \sin\th \, d\eta \wedge d\chi \wedge d\th \wedge d\phi \wedge d\psi \, .
\end{equation}
Notice that when $\m = 0$ one recovers the $AdS_5 \times S^5$ solution.

%%%%%%%%%%%%%%%%%%%%%%%%%%%%%%%%%%%%%%%%%%%%%%%%%%%%%%%%%%%

\section{String action}
\label{String_EOM}

We start with the (generalized) Polyakov action of a string propagating in a background with metric $G_{MN}$, NS two-form $B_{MN}$ and dilaton $\Phi$, which is:
\begin{equation}
 S = - \frac{1}{4 \pi \a'} \int d^2\s \sqrt{|h|} \Big(  h^{ab} \partial_a X^M \partial_b X^N G_{MN} 
 - \varepsilon^{ab} \partial_a X^M \partial_b X^N B_{MN}
 + \a' R^{(h)} \Phi \Big) \, ,
\end{equation}
where $h_{ab}$ is the world-sheet metric, $R^{(h)}$ is the Ricci scalar constructed with $h_{ab}$ and $\s^a = (\s^0 , \s^1) = (\tau , \s)$ are the string coordinates. Notice that $\varepsilon^{ab}$ is the Levi-Civita tensor, to be distinguished from the Levi-Civita symbol $\e^{ab}$ with $\e^{01} = 1$.

Things can get simpler going to the conformal gauge where:
\begin{equation}
 h_{ab} = \eta_{ab} = \textrm{diag}(-1 , 1) \, .
\end{equation}
Then the string action reads:
\begin{equation}
 S = - \frac{1}{4 \pi \a'} \int d^2\s \Big(  \eta^{ab} \partial_a X^M \partial_b X^N G_{MN} 
 + \e^{ab} \partial_a X^M \partial_b X^N B_{MN} \Big) \, .
\end{equation}
%
%The equations of motion for the string are:
%
%\begin{equation}
 %2 G_{MN} \Box X^N  - \eta^{ab} \partial_a X^P %\partial_b X^\S \partial_M G_{P\S} 
% - \e^{ab} \partial_a X^P \partial_b X^\S \partial_M B_{P\S} = 0 \, ,
%\end{equation}
%
%where $\Box = \eta^{ab} \partial_a \partial_b$. 
This has to be supplemented by the Virasoro constraints which at the conformal gauge take the form:
%
%\begin{equation}
% \partial_a X^M \partial_b X^N G_{MN} - \frac{1}{2} \eta_{ab} \partial X^M \cdot \partial X^N G_{MN} = 0 \, ,
%\end{equation}
%
%where the dot product refers to contractions with respect to $\eta_{ab}$. The last leads to:
%
\begin{equation}
 \Big(  \partial_\tau X^M \partial_\tau X^N + \partial_\s X^M \partial_\s X^N  \Big) G_{MN} = 0 \, , \qquad \partial_\tau X^M \partial_\s X^N G_{MN} = 0 \, .
\end{equation}

%%%%%%%%%%%%%%%%%%%%%%%%%%%%%%%%%%%%%%%%%%%%%%%%%%%%%%%% 

%%%%%%%%%%%%%%%%%%%%%%%%%%%%%%%%%%%%%%%%%%%%%%%%%%%%%%%%%%%

%\bibliographystyle{utphys}
%\bibliography{BibItsios.bib}

\begin{thebibliography}{99}


%\cite{Maldacena:1997re}
\bibitem{Maldacena:1997re}
  J.~M.~Maldacena,
  {\it The Large N limit of superconformal field theories and supergravity},
  Int.\ J.\ Theor.\ Phys.\  {\bf 38} (1999) 1113,
   [Adv.\ Theor.\ Math.\ Phys.\  {\bf 2} (1998) 231]
  %doi:10.1023/A:1026654312961, 10.4310/ATMP.1998.v2.n2.a1
  \href{https://arxiv.org/abs/hep-th/9711200}{{\tt hep-th/9711200}}.
  %%CITATION = doi:10.1023/A:1026654312961, 10.4310/ATMP.1998.v2.n2.a1;%%
  %13721 citations counted in INSPIRE as of 01 Jun 2018
  
  
%\cite{Staudacher:2004tk}
\bibitem{Staudacher:2004tk} 
  M.~Staudacher,
  {\it The Factorized S-matrix of CFT/AdS},
  JHEP {\bf 0505}, 054 (2005),
 % doi:10.1088/1126-6708/2005/05/054
  \href{https://arxiv.org/abs/hep-th/0412188}{{\tt hep-th/0412188}}.
  %%CITATION = doi:10.1088/1126-6708/2005/05/054;%%
  
 %\cite{Ambjorn:2005wa}
\bibitem{Ambjorn:2005wa} 
  J.~Ambjorn, R.~A.~Janik and C.~Kristjansen,
  {\it Wrapping interactions and a new source of corrections to the spin-chain/string duality},
  Nucl.\ Phys.\ B {\bf 736}, 288 (2006),
  %doi:10.1016/j.nuclphysb.2005.12.007
  \href{https://arxiv.org/abs/hep-th/0510171}{{\tt hep-th/0510171}}.
  %%CITATION = doi:10.1016/j.nuclphysb.2005.12.007;%%

  %\cite{Gromov:2009tv}
\bibitem{Gromov:2009tv} 
  N.~Gromov, V.~Kazakov and P.~Vieira,
  {\it Exact Spectrum of Anomalous Dimensions of Planar N=4 Supersymmetric Yang-Mills Theory},
  Phys.\ Rev.\ Lett.\  {\bf 103}, 131601 (2009),
  %doi:10.1103/PhysRevLett.103.131601
  \href{https://arxiv.org/abs/0901.3753}{{\tt arXiv:0901.3753}}.
  %%CITATION = doi:10.1103/PhysRevLett.103.131601;%%


  %\cite{Georgiou:2008vk}
\bibitem{Georgiou:2008vk}
  G.~Georgiou, V.~L.~Gili and R.~Russo,
  {\it Operator Mixing and the AdS/CFT correspondence},
  JHEP {\bf 0901}, 082 (2009),
  %doi:10.1088/1126-6708/2009/01/082
  \href{https://arxiv.org/abs/0810.0499}{{\tt arXiv:0810.0499}}.
  %%CITATION = doi:10.1088/1126-6708/2009/01/082;%%
  %16 citations counted in INSPIRE as of 22 May 2018

  %\cite{Georgiou:2009tp}
\bibitem{Georgiou:2009tp}
  G.~Georgiou, V.~L.~Gili and R.~Russo,
  {\it Operator mixing and three-point functions in N=4 SYM},
  JHEP {\bf 0910}, 009 (2009),
 % doi:10.1088/1126-6708/2009/10/009
  \href{https://arxiv.org/abs/0907.1567}{{\tt arXiv:0907.1567}}.
  %%CITATION = doi:10.1088/1126-6708/2009/10/009;%%
  %25 citations counted in INSPIRE as of 22 May 2018

  %\cite{Georgiou:2011xj}
\bibitem{Georgiou:2011xj}
  G.~Georgiou, V.~Gili and J.~Plefka,
  {\it The two-loop dilatation operator of N=4 super Yang-Mills theory in the SO(6) sector},
  JHEP {\bf 1112}, 075 (2011),
  %doi:10.1007/JHEP12(2011)075
  \href{https://arxiv.org/abs/1106.0724}{{\tt arXiv:1106.0724}}.
  %%CITATION = doi:10.1007/JHEP12(2011)075;%%
  %7 citations counted in INSPIRE as of 22 May 2018

%\cite{Okuyama:2004bd}
\bibitem{Okuyama:2004bd}
  K.~Okuyama and L.~S.~Tseng,
  {\it Three-point functions in N = 4 SYM theory at one-loop},
  JHEP {\bf 0408}, 055 (2004),
  %doi:10.1088/1126-6708/2004/08/055
  \href{https://arxiv.org/abs/hep-th/0404190}{{\tt hep-th/0404190}}.
  %%CITATION = doi:10.1088/1126-6708/2004/08/055;%%
  %85 citations counted in INSPIRE as of 22 May 2018

%\cite{Roiban:2004va}
\bibitem{Roiban:2004va}
  R.~Roiban and A.~Volovich,
  {\it Yang-Mills correlation functions from integrable spin chains},
  JHEP {\bf 0409}, 032 (2004),
  %doi:10.1088/1126-6708/2004/09/032
  \href{https://arxiv.org/abs/hep-th/0407140}{{\tt hep-th/0407140}}.
  %%CITATION = doi:10.1088/1126-6708/2004/09/032;%%
  %79 citations counted in INSPIRE as of 22 May 2018

%\cite{Alday:2005nd}
\bibitem{Alday:2005nd}
  L.~F.~Alday, J.~R.~David, E.~Gava and K.~S.~Narain,
  {\it Structure constants of planar N = 4 Yang Mills at one loop},
  JHEP {\bf 0509}, 070 (2005),
  %doi:10.1088/1126-6708/2005/09/070
  \href{https://arxiv.org/abs/hep-th/0502186}{{\tt hep-th/0502186}}.
  %%CITATION = doi:10.1088/1126-6708/2005/09/070;%%

  %\cite{Georgiou:2012zj}
\bibitem{Georgiou:2012zj}
  G.~Georgiou, V.~Gili, A.~Grossardt and J.~Plefka,
  {\it Three-point functions in planar N=4 super Yang-Mills Theory for scalar operators up to length
   five at the one-loop order},
  JHEP {\bf 1204}, 038 (2012),
  % doi:10.1007/JHEP04(2012)038
  \href{https://arxiv.org/abs/1201.0992}{{\tt arXiv:1201.0992}}.
  %%CITATION = doi:10.1007/JHEP04(2012)038;%%
  %32 citations counted in INSPIRE as of 22 May 2018
  
  %\cite{Escobedo:2010xs}
\bibitem{Escobedo:2010xs}
  J.~Escobedo, N.~Gromov, A.~Sever and P.~Vieira,
  {\it Tailoring Three-Point Functions and Integrability},
  JHEP {\bf 1109}, 028 (2011),
  %doi:10.1007/JHEP09(2011)028
  \href{https://arxiv.org/abs/1012.2475}{{\tt arXiv:1012.2475}}.

  %\cite{Jiang:2014mja}
\bibitem{Jiang:2014mja}
  Y.~Jiang, I.~Kostov, F.~Loebbert and D.~Serban,
  {\it Fixing the Quantum Three-Point Function},
  JHEP {\bf 1404}, 019 (2014),
  \href{https://arxiv.org/abs/1401.0384}{{\tt arXiv:1401.0384}}.
  %%CITATION = ARXIV:1401.0384;%%
  %7 citations counted in INSPIRE as of 24 Apr 2015

  %\cite{Basso:2015zoa}
\bibitem{Basso:2015zoa}
  B.~Basso, S.~Komatsu and P.~Vieira,
  {\it Structure Constants and Integrable Bootstrap in Planar N=4 SYM Theory},
  \href{https://arxiv.org/abs/1505.06745}{{\tt arXiv:1505.06745}}.
  %%CITATION = ARXIV:1505.06745;%%
  %73 citations counted in INSPIRE as of 22 May 2018

  %\cite{Kazama:2016cfl}
\bibitem{Kazama:2016cfl}
  Y.~Kazama, S.~Komatsu and T.~Nishimura,
  {\it Classical Integrability for Three-point Functions: Cognate Structure at Weak and Strong Couplings},
  JHEP {\bf 1610}, 042 (2016),
  %doi:10.1007/JHEP10(2016)042
  \href{https://arxiv.org/abs/1603.03164}{{\tt arXiv:1603.03164}}.
  %%CITATION = doi:10.1007/JHEP10(2016)042;%%
  

%\cite{Kazama:2013qsa}
\bibitem{Kazama:2013qsa}
  Y.~Kazama and S.~Komatsu,
  {\it Three-point functions in the SU(2) sector at strong coupling},
  JHEP {\bf 1403}, 052 (2014),
  %doi:10.1007/JHEP03(2014)052
  \href{https://arxiv.org/abs/1312.3727}{{\tt arXiv:1312.3727}}.
  %%CITATION = doi:10.1007/JHEP03(2014)052;%%
  %33 citations counted in INSPIRE as of 22 May 2018

  %\cite{Kazama:2011cp}
\bibitem{Kazama:2011cp}
  Y.~Kazama and S.~Komatsu,
  {\it On holographic three point functions for GKP strings from integrability},
  JHEP {\bf 1201}, 110 (2012),
  %Erratum: [JHEP {\bf 1206}, 150 (2012)]
  %doi:10.1007/JHEP06(2012)150, 10.1007/JHEP01(2012)110
  \href{https://arxiv.org/abs/1110.3949}{{\tt arXiv:1110.3949}}.
  %%CITATION = doi:10.1007/JHEP06(2012)150, 10.1007/JHEP01(2012)110;%%
  %56 citations counted in INSPIRE as of 22 May 2018

   %\cite{Zarembo:2010rr}
\bibitem{Zarembo:2010rr}
  K.~Zarembo,
  {\it Holographic three-point functions of semiclassical states},
  JHEP {\bf 1009} (2010) 030,
  %doi:10.1007/JHEP09(2010)030
  \href{https://arxiv.org/abs/1008.1059}{{\tt arXiv:1008.1059}}.
  %%CITATION = doi:10.1007/JHEP09(2010)030;%%
  %122 citations counted in INSPIRE as of 22 Feb 2018


  %\cite{Costa:2010rz}
\bibitem{Costa:2010rz}
  M.~S.~Costa, R.~Monteiro, J.~E.~Santos and D.~Zoakos,
  {\it On three-point correlation functions in the gauge/gravity duality},
  JHEP {\bf 1011} (2010) 141,
  %doi:10.1007/JHEP11(2010)141
  \href{https://arxiv.org/abs/1008.1070}{{\tt arXiv:1008.1070}}.
  %%CITATION = doi:10.1007/JHEP11(2010)141;%%
  %123 citations counted in INSPIRE as of 22 Feb 2018

  %\cite{Roiban:2010fe}
\bibitem{Roiban:2010fe}
  R.~Roiban and A.~A.~Tseytlin,
  {\it On semiclassical computation of 3-point functions of closed string vertex operators in $AdS_5 x S^5$},
  Phys.\ Rev.\ D {\bf 82} (2010) 106011,
  %doi:10.1103/PhysRevD.82.106011
  \href{https://arxiv.org/abs/1008.4921}{{\tt arXiv:1008.4921}}.
  %%CITATION = doi:10.1103/PhysRevD.82.106011;%%
  %89 citations counted in INSPIRE as of 22 Feb 2018

  %\cite{Georgiou:2010an}
\bibitem{Georgiou:2010an}
  G.~Georgiou,
  {\it Two and three-point correlators of operators dual to folded string solutions at strong coupling},
  JHEP {\bf 1102}, 046 (2011),
  %doi:10.1007/JHEP02(2011)046
  \href{https://arxiv.org/abs/1011.5181}{{\tt arXiv:1011.5181}}.
  %%CITATION = doi:10.1007/JHEP02(2011)046;%%
  %55 citations counted in INSPIRE as of 22 May 2018

  %\cite{Georgiou:2011qk}
\bibitem{Georgiou:2011qk}
  G.~Georgiou,
  {\it SL(2) sector: weak/strong coupling agreement of three-point correlators},
  JHEP {\bf 1109}, 132 (2011),
 % doi:10.1007/JHEP09(2011)132
  \href{https://arxiv.org/abs/1107.1850}{{\tt arXiv:1107.1850}}.
  %%CITATION = doi:10.1007/JHEP09(2011)132;%%
  %46 citations counted in INSPIRE as of 22 May 2018

  %\cite{Bajnok:2016xxu}
\bibitem{Bajnok:2016xxu}
  Z.~Bajnok and R.~A.~Janik,
  {\it Classical limit of diagonal form factors and HHL correlators},
  JHEP {\bf 1701}, 063 (2017),
  %doi:10.1007/JHEP01(2017)063
  \href{https://arxiv.org/abs/1607.02830}{{\tt arXiv:1607.02830}}.
  %%CITATION = doi:10.1007/JHEP01(2017)063;%%
  %4 citations counted in INSPIRE as of 23 May 2018

  %\cite{Berenstein:2002jq}
\bibitem{Berenstein:2002jq}
  D.~E.~Berenstein, J.~M.~Maldacena and H.~S.~Nastase,
  {\it Strings in flat space and pp waves from N=4 superYang-Mills},
  JHEP {\bf 0204}, 013 (2002),
  %doi:10.1088/1126-6708/2002/04/013
  \href{https://arxiv.org/abs/hep-th/0202021}{{\tt hep-th/0202021}}.
  %%CITATION = doi:10.1088/1126-6708/2002/04/013;%%

  %\cite{Spradlin:2002ar}
\bibitem{Spradlin:2002ar}
  M.~Spradlin and A.~Volovich,
  {\it Superstring interactions in a p p wave background},
  Phys.\ Rev.\ D {\bf 66}, 086004 (2002),
  %doi:10.1103/PhysRevD.66.086004
  \href{https://arxiv.org/abs/hep-th/0204146}{{\tt hep-th/0204146}}.
  %%CITATION = doi:10.1103/PhysRevD.66.086004;%%
  %177 citations counted in INSPIRE as of 22 May 2018

  %\cite{Pankiewicz:2002tg}
\bibitem{Pankiewicz:2002tg}
  A.~Pankiewicz and B.~Stefanski, Jr.,
  {\it PP wave light cone superstring field theory},
  Nucl.\ Phys.\ B {\bf 657}, 79 (2003),
  %doi:10.1016/S0550-3213(03)00141-X
  \href{https://arxiv.org/abs/hep-th/0210246}{{\tt hep-th/0210246}}.
  %%CITATION = doi:10.1016/S0550-3213(03)00141-X;%%
  %103 citations counted in INSPIRE as of 22 May 2018

  %\cite{DiVecchia:2003yp}
\bibitem{DiVecchia:2003yp}
  P.~Di Vecchia, J.~L.~Petersen, M.~Petrini, R.~Russo and A.~Tanzini,
  {\it The Three string vertex and the AdS / CFT duality in the PP wave limit},
  Class.\ Quant.\ Grav.\  {\bf 21}, 2221 (2004),
  %doi:10.1088/0264-9381/21/9/001
  \href{https://arxiv.org/abs/hep-th/0304025}{{\tt hep-th/0304025}}.
  %%CITATION = doi:10.1088/0264-9381/21/9/001;%%
  %58 citations counted in INSPIRE as of 22 May 2018

  %\cite{Dobashi:2004nm}
\bibitem{Dobashi:2004nm}
  S.~Dobashi and T.~Yoneya,
  {\it Resolving the holography in the plane-wave limit of AdS/CFT correspondence},
  Nucl.\ Phys.\ B {\bf 711}, 3 (2005),
  % doi:10.1016/j.nuclphysb.2005.01.024
  \href{https://arxiv.org/abs/hep-th/0406225}{{\tt hep-th/0406225}}.
  %%CITATION = doi:10.1016/j.nuclphysb.2005.01.024;%%

  %\cite{Lee:2004cq}
\bibitem{Lee:2004cq}
  S.~Lee and R.~Russo,
  {\it Holographic cubic vertex in the pp-wave},
  Nucl.\ Phys.\ B {\bf 705}, 296 (2005),
  %doi:10.1016/j.nuclphysb.2004.10.052
  \href{https://arxiv.org/abs/hep-th/0409261}{{\tt hep-th/0409261}}.
  %%CITATION = doi:10.1016/j.nuclphysb.2004.10.052;%%

  %\cite{Georgiou:2004ty}
\bibitem{Georgiou:2004ty}
  G.~Georgiou and G.~Travaglini,
  {\it Fermion BMN operators, the dilatation operator of N=4 SYM, and pp wave string interactions},
  JHEP {\bf 0404}, 001 (2004),
  %doi:10.1088/1126-6708/2004/04/001
  \href{https://arxiv.org/abs/hep-th/0403188}{{\tt hep-th/0403188}}.
  %%CITATION = doi:10.1088/1126-6708/2004/04/001;%%
  %21 citations counted in INSPIRE as of 22 May 2018

  %\cite{Georgiou:2003kt}
\bibitem{Georgiou:2003kt}
  G.~Georgiou, V.~V.~Khoze and G.~Travaglini,
  {\it New tests of the pp wave correspondence},
  JHEP {\bf 0310}, 049 (2003),
  %doi:10.1088/1126-6708/2003/10/049
  \href{https://arxiv.org/abs/hep-th/0306234}{{\tt hep-th/0306234}}.
  %%CITATION = doi:10.1088/1126-6708/2003/10/049;%%
  %21 citations counted in INSPIRE as of 22 May 2018

  %\cite{Georgiou:2003aa}
\bibitem{Georgiou:2003aa}
  G.~Georgiou and V.~V.~Khoze,
  {\it BMN operators with three scalar impurites and the vertex correlator duality in pp wave},
  JHEP {\bf 0304}, 015 (2003),
 % doi:10.1088/1126-6708/2003/04/015
  \href{https://arxiv.org/abs/hep-th/0302064}{{\tt hep-th/0302064}}.
  %%CITATION = doi:10.1088/1126-6708/2003/04/015;%%
  %23 citations counted in INSPIRE as of 22 May 2018

  %\cite{Chu:2002pd}
\bibitem{Chu:2002pd}
  C.~S.~Chu, V.~V.~Khoze and G.~Travaglini,
  {\it Three point functions in N=4 Yang-Mills theory and pp waves},
  JHEP {\bf 0206}, 011 (2002),
  %doi:10.1088/1126-6708/2002/06/011
  \href{https://arxiv.org/abs/hep-th/0206005}{{\tt hep-th/0206005}}.
  %%CITATION = doi:10.1088/1126-6708/2002/06/011;%%
  %82 citations counted in INSPIRE as of 22 May 2018

  %\cite{Georgiou:2016kge}
\bibitem{Georgiou:2016kge}
  G.~Georgiou and D.~Zoakos,
  {\it Entanglement Entropy of the $ \mathcal{N}=4 $ SYM spin chain},
  JHEP {\bf 1606}, 099 (2016),
  %doi:10.1007/JHEP06(2016)099
  \href{https://arxiv.org/abs/1603.05929}{{\tt arXiv:1603.05929}}.

  

  %\cite{Maldacena:2008wh}
\bibitem{Maldacena:2008wh}
  J.~Maldacena, D.~Martelli and Y.~Tachikawa,
  {\it Comments on string theory backgrounds with non-relativistic conformal symmetry},
  JHEP {\bf 0810}, 072 (2008),
  %doi:10.1088/1126-6708/2008/10/072
  \href{https://arxiv.org/abs/0807.1100}{{\tt arXiv:0807.1100}}.
  %%CITATION = doi:10.1088/1126-6708/2008/10/072;%%
  %328 citations counted in INSPIRE as of 23 May 2018
  
  %\cite{Herzog:2008wg}
\bibitem{Herzog:2008wg} 
  C.~P.~Herzog, M.~Rangamani and S.~F.~Ross,
  {\it Heating up Galilean holography},
  JHEP {\bf 0811}, 080 (2008),
  %doi:10.1088/1126-6708/2008/11/080
  \href{https://arxiv.org/abs/0807.1099}{{\tt arXiv:0807.1099}}.
  %%CITATION = doi:10.1088/1126-6708/2008/11/080;%%
  %303 citations counted in INSPIRE as of 02 Aug 2019
  
  %\cite{Adams:2008wt}
\bibitem{Adams:2008wt} 
  A.~Adams, K.~Balasubramanian and J.~McGreevy,
  {\it Hot Spacetimes for Cold Atoms},
  JHEP {\bf 0811}, 059 (2008),
  %doi:10.1088/1126-6708/2008/11/059
  \href{https://arxiv.org/abs/0807.1111}{{\tt arXiv:0807.1111}}.
  %%CITATION = doi:10.1088/1126-6708/2008/11/059;%%
  %299 citations counted in INSPIRE as of 02 Aug 2019

  %\cite{Alishahiha:2003ru}
\bibitem{Alishahiha:2003ru}
  M.~Alishahiha and O.~J.~Ganor,
  {\it Twisted backgrounds, PP waves and nonlocal field theories},
  JHEP {\bf 0303}, 006 (2003),
  %doi:10.1088/1126-6708/2003/03/006
  \href{https://arxiv.org/abs/hep-th/0301080}{{\tt hep-th/0301080}}.
  %%CITATION = doi:10.1088/1126-6708/2003/03/006;%%
  %75 citations counted in INSPIRE as of 22 May 2018

  %\cite{Beisert:2005if}
\bibitem{Beisert:2005if}
  N.~Beisert and R.~Roiban,
  {\it Beauty and the twist: The Bethe ansatz for twisted N=4 SYM},
  JHEP {\bf 0508}, 039 (2005),
  %doi:10.1088/1126-6708/2005/08/039
  \href{https://arxiv.org/abs/hep-th/0505187}{{\tt hep-th/0505187}}.
  %%CITATION = doi:10.1088/1126-6708/2005/08/039;%%
  %155 citations counted in INSPIRE as of 22 May 2018

  %\cite{Ahn:2010ws}
\bibitem{Ahn:2010ws}
  C.~Ahn, Z.~Bajnok, D.~Bombardelli and R.~I.~Nepomechie,
  {\it Twisted Bethe equations from a twisted S-matrix},
  JHEP {\bf 1102}, 027 (2011),
 % doi:10.1007/JHEP02(2011)027
  \href{https://arxiv.org/abs/1010.3229}{{\tt arXiv:1010.3229}}.
  %%CITATION = doi:10.1007/JHEP02(2011)027;%%
  %38 citations counted in INSPIRE as of 22 May 2018
  
  %\cite{Matsumoto:2015uja}
\bibitem{Matsumoto:2015uja}
  T.~Matsumoto and K.~Yoshida,
  {\it Schrodinger geometries arising from Yang-Baxter deformations},
  JHEP {\bf 1504} (2015) 180,
  %doi:10.1007/JHEP04(2015)180
  \href{https://arxiv.org/abs/1502.00740}{{\tt arXiv:1502.00740}}.
  %%CITATION = doi:10.1007/JHEP04(2015)180;%%
  %42 citations counted in INSPIRE as of 01 Jul 2019
  
  %\cite{Kyono:2016jqy}
\bibitem{Kyono:2016jqy} 
  H.~Kyono and K.~Yoshida,
  {\it Supercoset construction of Yang-Baxter deformed AdS$_5\times$S$^5$ backgrounds},
  PTEP {\bf 2016}, no. 8, 083B03 (2016),
  %doi:10.1093/ptep/ptw111
  \href{https://arxiv.org/abs/1605.02519}{{\tt arXiv:1605.02519}}.
  %%CITATION = doi:10.1093/ptep/ptw111;%%
  %39 citations counted in INSPIRE as of 02 Aug 2019
  
  %\cite{vanTongeren:2015uha}
\bibitem{vanTongeren:2015uha}
  S.~J.~van Tongeren,
  {\it Yang--Baxter deformations, AdS/CFT, and twist-noncommutative gauge theory},
  Nucl.\ Phys.\ B {\bf 904} (2016) 148,
  %doi:10.1016/j.nuclphysb.2016.01.012
  \href{https://arxiv.org/abs/1506.01023}{{\tt arXiv:1506.01023}}.
  %%CITATION = doi:10.1016/j.nuclphysb.2016.01.012;%%
  %74 citations counted in INSPIRE as of 01 Jul 2019

  %\cite{Fuertes:2009ex}
\bibitem{Fuertes:2009ex}
  C.~A.~Fuertes and S.~Moroz,
  {\it Correlation functions in the non-relativistic AdS/CFT correspondence},
  Phys.\ Rev.\ D {\bf 79} (2009) 106004,
  %doi:10.1103/PhysRevD.79.106004
  \href{https://arxiv.org/abs/0903.1844}{{\tt arXiv:0903.1844}}.
  %%CITATION = doi:10.1103/PhysRevD.79.106004;%%
  %43 citations counted in INSPIRE as of 20 Feb 2018

   %\cite{Volovich:2009yh}
\bibitem{Volovich:2009yh}
  A.~Volovich and C.~Wen,
  {\it Correlation Functions in Non-Relativistic Holography},
  JHEP {\bf 0905} (2009) 087,
  %doi:10.1088/1126-6708/2009/05/087
  \href{https://arxiv.org/abs/0903.2455}{{\tt arXiv:0903.2455}}.
  %%CITATION = doi:10.1088/1126-6708/2009/05/087;%%
  %82 citations counted in INSPIRE as of 20 Feb 2018

 %\cite{Georgiou:2017pvi}
\bibitem{Georgiou:2017pvi}
  G.~Georgiou and D.~Zoakos,
  {\it Giant magnons and spiky strings in the Schrodinger/ dipole-deformed CFT correspondence},
  JHEP {\bf 1802} (2018) 173,
  %doi:10.1007/JHEP02(2018)173
  \href{https://arxiv.org/abs/1712.03091}{{\tt arXiv:1712.03091}}.
  %%CITATION = doi:10.1007/JHEP02(2018)173;%%
  %1 citations counted in INSPIRE as of 21 Mar 2018
  
  
  %\cite{Zoakos:2020gyb}
\bibitem{Zoakos:2020gyb}
D.~Zoakos,
{\it Finite size effects in classical string solutions of the Schr\"odinger geometry},
JHEP \textbf{08}, 091 (2020),
%doi:10.1007/JHEP08(2020)091
\href{https://arxiv.org/abs/2006.02285}{{\tt arXiv:2006.02285}}.
%4 citations counted in INSPIRE as of 04 Jul 2022


%\cite{Georgiou:2019lqh}
\bibitem{Georgiou:2019lqh}
G.~Georgiou, K.~Sfetsos and D.~Zoakos,
{\it String theory on the Schr\"odinger pp-wave background},
JHEP \textbf{08} (2019), 093,
%doi:10.1007/JHEP08(2019)093
\href{https://arxiv.org/abs/1906.08269}{{\tt arXiv:1906.08269}}.
%6 citations counted in INSPIRE as of 30 Apr 2022


%\cite{Georgiou:2018zkt}
\bibitem{Georgiou:2018zkt}
  G.~Georgiou and D.~Zoakos,
  {\it Holographic three-point correlators in the Schrodinger/dipole CFT correspondence},
  JHEP {\bf 1809}, 026 (2018),
  %doi:10.1007/JHEP09(2018)026
  \href{https://arxiv.org/abs/1806.08181}{{\tt arXiv:1806.08181}}.
  %%CITATION = doi:10.1007/JHEP09(2018)026;%%

%\cite{Guica:2017mtd}
\bibitem{Guica:2017mtd}
M.~Guica, F.~Levkovich-Maslyuk and K.~Zarembo,
{\it Integrability in dipole-deformed $\boldsymbol{\mathcal{N}=4}$ super Yang\textendash{}Mills},
J. Phys. A \textbf{50} (2017) no.39, 39,
%doi:10.1088/1751-8121/aa8491
\href{https://arxiv.org/abs/1706.07957}{{\tt arXiv:1706.07957}}.
%31 citations counted in INSPIRE as of 19 Feb 2021

  %\cite{Ouyang:2017yko}
\bibitem{Ouyang:2017yko}
  H.~Ouyang,
  {\it Semiclassical spectrum for BMN string in $Sch_5\times S^5$},
  JHEP {\bf 1712}, 126 (2017),
  %doi:10.1007/JHEP12(2017)126
  \href{https://arxiv.org/abs/1709.06844}{{\tt arXiv:1709.06844}}.
  %%CITATION = doi:10.1007/JHEP12(2017)126;%%
  %2 citations counted in INSPIRE as of 23 May 2018


   %\cite{Dimov:2019koi}
\bibitem{Dimov:2019koi}
H.~Dimov, M.~Radomirov, R.~C.~Rashkov and T.~Vetsov,
{\it On pulsating strings in Schr\"odinger backgrounds},
JHEP \textbf{10}, 094 (2019),
%doi:10.1007/JHEP10(2019)094
\href{https://arxiv.org/abs/1903.07444}{{\tt arXiv:1903.07444}}.
%7 citations counted in INSPIRE as of 12 Jul 2022
  
  %\cite{Georgiou:2020wwo}
\bibitem{Georgiou:2020wwo}
G.~Georgiou and D.~Zoakos,
{\it Giant gravitons in the Schr\"odinger holography},
JHEP \textbf{01}, 017 (2021),
%doi:10.1007/JHEP01(2021)017
\href{https://arxiv.org/abs/2010.03952}{{\tt arXiv:2010.03952}}.
%0 citations counted in INSPIRE as of 24 Jun 2022

%\cite{Georgiou:2020qnh}
\bibitem{Georgiou:2020qnh}
G.~Georgiou and D.~Zoakos,
{\it Giant gravitons on the Schr\"odinger pp-wave geometry},
JHEP \textbf{03} (2020), 185,
%doi:10.1007/JHEP03(2020)185
\href{https://arxiv.org/abs/2002.05460}{{\tt arXiv:2002.05460}}.
%5 citations counted in INSPIRE as of 30 Apr 2022



%\cite{Pirrone:2006iq}
\bibitem{Pirrone:2006iq}
M.~Pirrone,
{\it Giants On Deformed Backgrounds},
JHEP \textbf{12}, 064 (2006),
%doi:10.1088/1126-6708/2006/12/064
\href{https://arxiv.org/abs/hep-th/0609173}{{\tt arXiv:hep-th/0609173}}.
%37 citations counted in INSPIRE as of 24 Jun 2022


%\cite{Imeroni:2006rb}
\bibitem{Imeroni:2006rb}
E.~Imeroni and A.~Naqvi,
{\it Giants and loops in beta-deformed theories},
JHEP \textbf{03}, 034 (2007),
%doi:10.1088/1126-6708/2007/03/034
\href{https://arxiv.org/abs/hep-th/0612032}{{\tt arXiv:hep-th/0612032}}.
%28 citations counted in INSPIRE as of 24 Jun 2022


%\cite{Maldacena:1998im}
\bibitem{Maldacena:1998im}
J.~M.~Maldacena,
{\it Wilson loops in large N field theories},
Phys. Rev. Lett. \textbf{80}, 4859-4862 (1998),
%doi:10.1103/PhysRevLett.80.4859
\href{https://arxiv.org/abs/hep-th/9803002}{{\tt arXiv:hep-th/9803002}}.
%1834 citations counted in INSPIRE as of 25 May 2022

%\cite{Armas:2014nea}
\bibitem{Armas:2014nea}
J.~Armas and M.~Blau,
{\it Black probes of Schr\"odinger spacetimes},
JHEP \textbf{08}, 140 (2014),
%doi:10.1007/JHEP08(2014)140
\href{https://arxiv.org/abs/1405.1301}{{\tt arXiv:1405.1301}}.
%10 citations counted in INSPIRE as of 26 May 2022


%\cite{Araujo:2015qga}
\bibitem{Araujo:2015qga}
T.~R.~Araujo,
{\it Revisiting Wilson loops for nonrelativistic backgrounds},
Phys. Rev. D \textbf{92}, no.12, 126007 (2015),
%doi:10.1103/PhysRevD.92.126007
\href{https://arxiv.org/abs/1509.02011}{{\tt arXiv:1509.02011}}.
%1 citations counted in INSPIRE as of 26 May 2022


%\cite{Dasgupta:2001zu}
\bibitem{Dasgupta:2001zu}
K.~Dasgupta and M.~M.~Sheikh-Jabbari,
{\it Noncommutative dipole field theories},
JHEP \textbf{02}, 002 (2002),
%doi:10.1088/1126-6708/2002/02/002
\href{https://arxiv.org/abs/hep-th/0112064}{{\tt arXiv:hep-th/0112064}}.
%58 citations counted in INSPIRE as of 14 Jun 2022

\bibitem{Kruczenski:2003gt}
M.~Kruczenski,
{\it Spin chains and string theory},
Phys. Rev. Lett. \textbf{93}, 161602 (2004),
%doi:10.1103/PhysRevLett.93.161602
\href{https://arxiv.org/abs/hep-th/0311203}{{\tt arXiv:hep-th/0311203}}.

%\cite{Akhavan:2008ep}
\bibitem{Akhavan:2008ep}
A.~Akhavan, M.~Alishahiha, A.~Davody and A.~Vahedi,
{\it Non-relativistic CFT and Semi-classical Strings},
JHEP \textbf{03}, 053 (2009),
%doi:10.1088/1126-6708/2009/03/053
\href{https://arxiv.org/abs/0811.3067}{{\tt arXiv:0811.3067}}.
%37 citations counted in INSPIRE as of 26 May 2022

%\cite{Hernandez:2005xd}
\bibitem{Hernandez:2005xd}
R.~Hernandez, K.~Sfetsos and D.~Zoakos,
{\it Gravity duals for the Coulomb branch of marginally deformed ${\cal N}=4$ Yang-Mills},
JHEP \textbf{03}, 069 (2006),
%doi:10.1088/1126-6708/2006/03/069
\href{https://arxiv.org/abs/hep-th/0510132}{{\tt arXiv:hep-th/0510132}}.
%35 citations counted in INSPIRE as of 26 May 2022

%\cite{Hernandez:2005zx}
\bibitem{Hernandez:2005zx}
R.~Hernandez, K.~Sfetsos and D.~Zoakos,
{\it On supersymmetry and other properties of a class of marginally deformed backgrounds},
Fortsch. Phys. \textbf{54}, 407-415 (2006),
%doi:10.1002/prop.200510294
\href{https://arxiv.org/abs/hep-th/0512158}{{\tt arXiv:hep-th/0512158}}.
%10 citations counted in INSPIRE as of 27 May 2022


%\cite{Pirrone:2008av}
\bibitem{Pirrone:2008av}
M.~Pirrone,
{\it Giants on Deformed Backgrounds Part II: The Gauge Field Fluctuations},
JHEP \textbf{03}, 034 (2008),
%doi:10.1088/1126-6708/2008/03/034
\href{https://arxiv.org/abs/0801.2540}{{\tt arXiv:0801.2540}}.
%10 citations counted in INSPIRE as of 24 Jun 2022


%\cite{Grisaru:2000zn}
\bibitem{Grisaru:2000zn}
M.~T.~Grisaru, R.~C.~Myers and O.~Tafjord,
{\it SUSY and goliath},
JHEP \textbf{08}, 040 (2000),
%doi:10.1088/1126-6708/2000/08/040
\href{https://arxiv.org/abs/hep-th/0008015}{{\tt arXiv:hep-th/0008015}}.
%366 citations counted in INSPIRE as of 24 Jun 2022


%\cite{Niarchos:2002fc}
\bibitem{Niarchos:2002fc}
V.~Niarchos and N.~Prezas,
{\it BMN operators for N=1 superconformal Yang-Mills theories and associated string backgrounds},
JHEP \textbf{06}, 015 (2003),
%doi:10.1088/1126-6708/2003/06/015
\href{https://arxiv.org/abs/hep-th/0212111}{{\tt arXiv:hep-th/0212111}}.
%50 citations counted in INSPIRE as of 27 Jun 2022

%\cite{Lunin:2005jy}
\bibitem{Lunin:2005jy}
O.~Lunin and J.~M.~Maldacena,
{\it Deforming field theories with U(1) x U(1) global symmetry and their gravity duals},
JHEP \textbf{05}, 033 (2005),
%doi:10.1088/1126-6708/2005/05/033
\href{https://arxiv.org/abs/hep-th/0502086}{{\tt arXiv:hep-th/0502086}}.
%632 citations counted in INSPIRE as of 27 Jun 2022

%\cite{Avramis:2007wb}
\bibitem{Avramis:2007wb}
S.~D.~Avramis, K.~Sfetsos and D.~Zoakos,
{\it Complex marginal deformations of D3-brane geometries, their Penrose limits and giant gravitons},
Nucl. Phys. B \textbf{787}, 55-97 (2007),
%doi:10.1016/j.nuclphysb.2007.07.017
\href{https://arxiv.org/abs/0704.2067}{{\tt arXiv:0704.2067}}.
%23 citations counted in INSPIRE as of 24 Jun 2022


%\cite{Chu:2006ae}
\bibitem{Chu:2006ae}
C.~S.~Chu, G.~Georgiou and V.~V.~Khoze,
{\it Magnons, classical strings and beta-deformations},
JHEP \textbf{11}, 093 (2006),
%doi:10.1088/1126-6708/2006/11/093
\href{https://arxiv.org/abs/hep-th/0606220}{{\tt arXiv:hep-th/0606220}}.




  
\end{thebibliography}

\end{document}